\documentclass[prd,floatfix,onecolumn,amsmath,amssymb,floatfix]{revtex4}
\usepackage{graphicx,color,dcolumn,booktabs,bm}
\usepackage{subfigure}
\usepackage{longtable,lscape}
\usepackage{amssymb}
\usepackage{indentfirst}
\usepackage{epsfig}
\usepackage{feynmf}   %{feynmp}
\usepackage{epstopdf}   %{feynmp}
\usepackage{slashed}  %for Feynman symbols
\usepackage{cases}
\usepackage[pdfpages]{xcolor}
\definecolor{maroon}{RGB}{139,25,150}%burada 0-255 arasi her biri icin numara vererek renk elde et
\usepackage{multirow}
\usepackage{float}
\usepackage{graphicx,color,dcolumn,booktabs,bm}
\usepackage[colorlinks, citecolor=blue,anchorcolor=red,menucolor=red, linkcolor=red,filecolor=red,runcolor=red,urlcolor=blue,frenchlinks=red, urlcolor=blue]{hyperref}
\usepackage{orcidlink}

\allowdisplaybreaks
\begin{document}
%%%%%%%%%%%%%%%%%%%%%%%%%%%%%%%%%

\preprint{}
\preprint{}
\title{\color{blue}{Investigation of the semileptonic decay $ \Xi^{++}_{cc}\rightarrow \Xi^+_{c} \bar{\ell}\nu_{\ell}$  within QCD sum rules}}

\author{M.~Shekari Tousi$^{a}$\orcidlink{0009-0007-7195-0838}}

	\author{K.~Azizi$^{a,b}$\orcidlink{0000-0003-3741-2167}} 
	\email{kazem.azizi@ut.ac.ir} \thanks{Corresponding author} 
	
	\author{H. R. Moshfegh$^{a,c}$\,\orcidlink{0000-0002-9657-7116}}

\affiliation{
		$^{a}$Department of Physics, University of Tehran, North Karegar Avenue, Tehran 14395-547, Iran\\
		$^{b}$Department of Physics, Do\v{g}u\c{s} University, Dudullu-\"{U}mraniye, 34775 Istanbul, T\"urkiye\\
        $^{c}$Centro Brasileiro de Pesquisas F´ısicas, Rua Dr. Xavier Sigaud,150, URCA, Rio de Janeiro CEP 22290-180, RJ, Brazil}

\date{\today}

\begin{abstract}

We study the semileptonic decay of the doubly heavy baryon $ \Xi^{++}_{cc} $ into the singly heavy baryon $ \Xi^+_{c}$ within the three-point QCD sum rule approach in two possible lepton channels. Our analysis includes perturbative as well as  nonperturbative condensation contributions up to dimension 5. We evaluate the form factors of this semileptonic decay entering the amplitude described by the  vector and axial vector transition currents.  The fit functions of the  form factors with respect to the transferred momentum squared are utilized to predict the decay widths and branching ratios of the $ \Xi^{++}_{cc}\rightarrow \Xi^+_{c} \bar{\ell}\nu_{\ell}$  channels. We compare our findings with other predictions in the literature. Our outcomes can be useful for experimental groups in their search for the weak decays of doubly heavy baryons and may be checked via future experiments such as LHCb.

\end{abstract}

\maketitle
\section{Introduction}
Recent research on hadrons containing heavy quarks has gained significant consideration. Despite the quark model's \cite{GellMann:1964nj} notable achievements in hadron spectroscopy, not all the predicted particles, including the doubly heavy baryons, have been experimentally confirmed to date. The first discovery of a doubly heavy baryon was reported in 2002, when the SELEX Collaboration announced the first observation of $\Xi^+_{cc}(3520)$ in the $ p D^+ K^-$ decay channel \cite{SELEX:2002wqn}. This discovery was further validated by the same Collaboration in 2005 \cite{SELEX:2004lln}. In a milestone, the LHCb Collaboration in 2017 reported the first experimental validation of a doubly charmed baryon $ \Xi^{++}_{cc} (3621)$ via the $ \Xi^{++}_{cc}\rightarrow \Lambda^+_{c} K^- \pi^+ \pi^+$ decay channel \cite{LHCb:2017iph}. This was later confirmed in 2018 through the $ \Xi^{++}_{cc}\rightarrow \Xi^+_{c} \pi^+$ decay channel \cite{LHCb:2018pcs}. 
In 2019, the LHCb Collaboration conducted a search for $\Xi^+_{cc}$ via the $ \Xi^{+}_{cc}\rightarrow \Lambda^+_{c} K^- \pi^+$ decay channel \cite{LHCb:2019gqy}. The results from this search were combined with results from another study conducted in 2021 on the decay channel \( \Xi^{+}_{cc}\rightarrow \Xi^{+}_{c} \pi^- \pi^+ \). This combination yielded a maximum local significance of 4.0 standard deviations near the mass of 3620 MeV for the \( \Xi^{+}_{cc} \), including systematic uncertainties \cite{LHCb:2021eaf}. This result does not represent the previous big difference  between the masses of $\Xi^{+}_{cc}$ and $ \Xi_{cc}^{++} $. Experimental efforts have continued to discover additional members of these baryon types. However, as of now, these efforts have not resulted in the discovery of any new doubly heavy baryons (see Ref. \cite{LHCb:2021xba}). Numerous particle physicists have performed research based on these experimental efforts to determine the features of the doubly heavy baryons such as their mass and residue \cite{ShekariTousi:2024mso, Ebert:2002ig,Zhang:2008rt,Wang:2010hs,Lu:2017meb,Rahmani:2020pol,Yao:2018ifh,Aliyev:2022rrf,Aliev:2012iv,Aliev:2019lvd,Aliev:2012ru,Padmanath:2019ybu,Brown:2014ena,Giannuzzi:2009gh,Shah:2017liu,Shah:2016vmd,Yoshida:2015tia,Valcarce:2008dr,Wang:2010it,Ortiz-Pacheco:2023kjn,Wang:2018lhz}, mixing angle \cite{Aliev:2012nn}, chiral effective Lagrangian \cite{Qiu:2020omj}, strong coupling constants \cite{Olamaei:2021hjd,Aliev:2021hqq,Aliev:2020lly,Alrebdi:2020rev,Rostami:2020euc,Olamaei:2020bvw,Aliev:2020aon}, strong interaction and decay \cite{Azizi:2020zin,Qin:2021dqo,Xiao:2017dly}, radiative decays \cite{Aliev:2021hqq,Xiao:2017udy,Lu:2017meb,Rahmani:2020pol,Li:2017pxa,Ortiz-Pacheco:2023kjn}, weak decays \cite{Gerasimov:2019jwp,Wang:2017mqp,Zhao:2018mrg,Xing:2018lre,Jiang:2018oak,Gutsche:2019wgu,Gutsche:2019iac,Ke:2019lcf,Cheng:2020wmk,Hu:2020mxk,Li:2020qrh,Han:2021gkl,Wang:2017azm,Shi:2017dto,Zhang:2018llc,Ivanov:2020xmw,Shi:2020qde,Hu:2017dzi,Li:2018epz,Shi:2019hbf,Shi:2019fph,Sharma:2017txj,Patel:2024mfn,Gutsche:2017hux,Gutsche:2018msz}, magnetic moments \cite{Ozdem:2018uue,Ozdem:2019zis}, lifetime \cite{Berezhnoy:2018bde}, etc., using different approaches. To calculate these parameters, we require nonperturbative approaches like the QCD sum rules, introduced in 1979 by Shifman, Vainshtein, and Zakharov \cite{Shifman:1978bx, Shifman:1978by}. This method is grounded in the fundamental QCD Lagrangian. In this nonperturbative approach, we consider the correlation function including various interpolating currents. This method has many successful predictions in the discussion of hadronic parameters confirmed by various experiments and is a very prosperous method  \cite{Aliev:2010uy,Aliev:2009jt,Aliev:2012ru,Agaev:2016dev,Azizi:2016dhy,Wang:2007ys}.

In this paper, we study the semileptonic decay of the doubly heavy baryon $ \Xi^{++}_{cc} $ into the singly heavy baryon $ \Xi^+_{c}$ in the framework of the QCD sum rule approach in two lepton channels. To investigate the semileptonic decay $ \Xi^{++}_{cc}\rightarrow \Xi^+_{c} \bar{\ell}\nu_{\ell}$, we need to calculate the form factors and after calculating them, we can achieve the decay widths and branching ratio of this decay.  In previous research, this transition has also been studied in QCD sum rule method  with different interpolating currents \cite{Shi:2019hbf} and light front approach \cite{Wang:2017mqp}, treating the two possible  leptons ($e^+$ and $\mu ^+$) the same.  In this study, we calculate the corresponding form factors by considering the most general interpolating  currents in the initial and final baryonic channels by utilizing the  recently calculated value of residue for  $ \Xi^{++}_{cc} $ \cite{ShekariTousi:2024mso}.  We estimate the decay widths and branching ratios  for the two lepton channels separately. By accurate fixing of the corresponding  parameters such as the mixing parameter in the interpolating currents, we will see that we can reach to the results with relatively small uncertainties.

This study is organized as follows: section \ref{Sec2} presents a brief derivation of sum rules to calculate the form factors of the semileptonic decay of $ \Xi^{++}_{cc}\rightarrow \Xi^+_{c} \bar{\ell}\nu_{\ell}$. Sec. \ref{Sec3} presents the findings from numerical analysis of the sum rules, showing fit functions that describe the behavior of form factors  as a function of transferred momentum square.  In section \ref{Sec4}, the calculation and results of the decay widths and branching ratios are shown for two possible lepton channels, followed by a comparison of our findings with other theoretical predictions. Section \ref{Sec5} contains the conclusion, while certain calculation details  are  provided in the Appendix.

\section{CALCULATION OF THE FORM FACTORS WITHIN THE QCD SUM RULES}~\label{Sec2}
 
The QCD sum rule follows a general procedure that involves evaluating a correlation function through two different approaches. The first approach utilizes hadronic degrees of freedom, known as the physical or phenomenological side, providing outcomes containing physical quantities such as the mass and residue of hadronic states. The second approach employs QCD degrees of freedom, including QCD coupling constants, quark-gluon condensates and quark masses, yielding the QCD side. By matching the outcomes of both approaches and considering the coefficients of the same Lorentz structures, QCD sum rules for the physical quantities are obtained.

\subsection{Phenomenological side}
The $ \Xi^{++}_{cc}\rightarrow \Xi^+_{c} \bar{\ell}\nu_{\ell}$ decay channel occurs via $c\to s \bar{\ell}\nu_{\ell}$ transition at quark level (see Fig. \ref{Fig:curre}):

\begin{figure}[h!] 
	\includegraphics[totalheight=2cm,width=6cm]{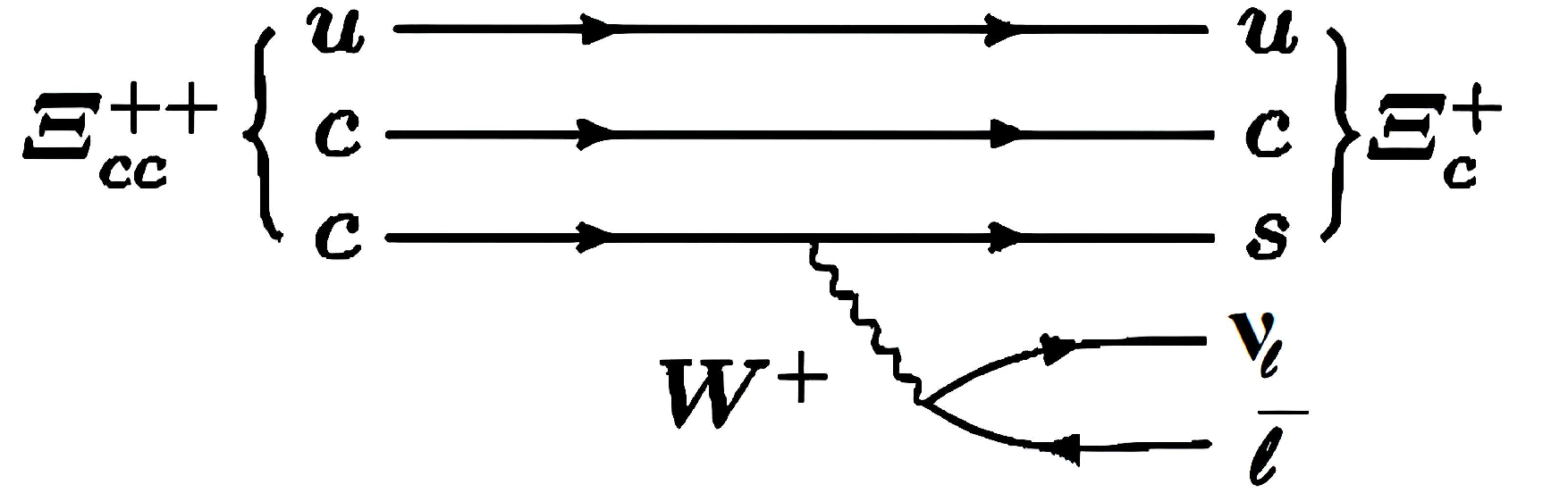}
	\caption{The $ \Xi^{++}_{cc}\rightarrow \Xi^+_{c} \bar{\ell}\nu_{\ell}$ decay channel.}\label{Fig:curre}
\end{figure}
To calculate the amplitude of this decay channel, we demand the low-energy effective Hamiltonian at the quark level. The transition current of this semileptonic decay can be written as:

 \begin{eqnarray}\label{jeff}
  J^{tr}=\bar s ~\gamma_\mu(1-\gamma_5)~ c. 
\end{eqnarray}
The effective Hamiltonian can be defined as: 
\begin{eqnarray}\label{Heff}
{\cal H}_{eff} =
\frac{G_F}{\sqrt2} V_{sc}^{*} ~\bar s \gamma_\mu(1-\gamma_5) c ~\bar\nu_{\ell}\gamma^\mu(1-\gamma_5){\ell} ,
\end{eqnarray}
where $V_{sc}^{*}$ stands for  the Cabibbo-Kobayashi-Maskawa (CKM) matrix elements and $G_F$ is the Fermi coupling constant. The decay amplitude is obtained by sandwiching the effective Hamiltonian between the initial  $ \vert  \Xi^{++}_{cc}\rangle$ and final $ \vert  \Xi^{+}_{c}\rangle$ states:

\begin{eqnarray}\label{amp}
&&M=\langle  \Xi^{+}_{c}\vert{\cal H}_{eff}\vert \Xi^{++}_{cc}\rangle \notag \\
&&=\frac{G_F}{\sqrt2} V_{sc}^{*} \bar\nu_{\ell}~\gamma^\mu(1-\gamma_5) {\ell} \langle  \Xi^{+}_{c}\vert \bar s \gamma_\mu(1-\gamma_5) c \vert  \Xi^{++}_{cc}\rangle.
\end{eqnarray}
This decay consists of the vector transition ($V^\mu$) and the axial vector transition ($A^\mu$), each of which is parametrized in terms of three form factors. The parametrizations, taking into account Lorentz invariance and parity considerations, are shown as \cite{Azizi:2018axf}: 

\begin{eqnarray}\label{Cur.with FormFac.}
&&\langle  \Xi^{+}_{c}(p',s')|V^{\mu}| \Xi^{++}_{cc} (p,s)\rangle = \bar
u_{ \Xi^{+}_{c}}(p',s') \Big[F_1(q^2)\gamma^{\mu}+F_2(q^2)\frac{p^{\mu}}{M_{ \Xi^{++}_{cc}}}
+F_3(q^2)\frac{p'^{\mu}}{M_{ \Xi^{+}_{c}}}\Big] u_{ \Xi^{++}_{cc}}(p,s), \notag \\
&&\langle  \Xi^{+}_{c}(p',s')|A^{\mu}| \Xi^{++}_{cc} (p,s)\rangle = \bar u_{ \Xi^{+}_{c}}(p',s') \Big[G_1(q^2)\gamma^{\mu}+G_2(q^2)\frac{p^{\mu}}{M_{ \Xi^{++}_{cc}}}+G_3(q^2)\frac{p'^{\mu}}{M_{ \Xi^{+}_{c}}}\Big]
\gamma_5 u_{ \Xi^{++}_{cc}}(p,s), \notag \\
\end{eqnarray}
where $ F_1(q^2),~F_2(q^2)$ and $ F_3(q^2)$ are form factors of the vector transition and  $ G_1(q^2),~G_2(q^2)$ and $G_3(q^2)$ are form factors corresponding to the axial transition. $u_{ \Xi^{++}_{cc}}(p,s)$ and
$u_{ \Xi^{+}_{c}}(p',s')$ are Dirac spinors of the initial and final baryon states and the momentum transferred to the leptons is $q=p-p'$. To calculate the form factors, we first utilize the three-point correlation function:
\begin{eqnarray}\label{CorFunc}
\Pi_{\mu}(p,p^{\prime},q)&=&i^2\int d^{4}x e^{-ip\cdot x}\int d^{4}y e^{ip'\cdot y}  \langle 0|{\cal T}\{{\cal J}^{\Xi^{+}_{c}}(y){\cal
J}_{\mu}^{tr,V(A)}(0) \bar {\cal J}^{  \Xi^{++}_{cc}}(x)\}|0\rangle,
\end{eqnarray}
where $ {\cal J}^{ \Xi^{++}_{cc}}(x)$ and ${\cal J}^{\Xi^{+}_{c}}(y) $ are the interpolating currents of the initial and final baryons  and  $\cal T$ is the time ordering operator. 
To calculate the three-point correlation function on the phenomenological side, it is necessary to insert two appropriate complete sets of hadronic states with identical quantum numbers as the currents ${\cal J}^{ \Xi^{++}_{cc}}$ and ${\cal J}^{\Xi^{+}_{c}}$ for both the initial and final baryons in the relevant places. By performing some algebraic calculations, the correlation function of the phenomenological side becomes: 
\begin{eqnarray} \label{PhysSide}
\Pi_{\mu}^{Phys.}(p,p',q)=\frac{\langle 0 \mid {\cal J}^{\Xi^{+}_{c}} (0)\mid \Xi^{+}_{c}(p') \rangle \langle \Xi^{+}_{c} (p')\mid
{\cal J}_{\mu}^{tr,V(A)}(0)\mid  \Xi^{++}_{cc}(p) \rangle \langle  \Xi^{++}_{cc}(p)
\mid \bar {\cal J}^{  \Xi^{++}_{cc}}(0)\mid
0\rangle}{(p'^2-m_{\Xi^{+}_{c}}^2)(p^2-m_{ \Xi^{++}_{cc}}^2)}+\cdots~,
\end{eqnarray}
where $\cdots$ represents the higher states and continuum contributions.
 We also define the residues of the initial ($\lambda_{ \Xi^{++}_{cc}}$) and final ($\lambda_{\Xi^{+}_{c}}$) baryons as follows:
\begin{eqnarray}\label{MatrixElements}
&&\langle 0|{\cal J}^{\Xi^{+}_{c}}(0)|\Xi^{+}_{c}(p')\rangle =
\lambda_{\Xi^{+}_{c}} u_{\Xi^{+}_{c}}(p',s'), \notag \\
&&\langle \Xi^{++}_{cc}(p)|\bar {\cal J}^{ \Xi^{++}_{cc}}(0)| 0 \rangle =
\lambda^{+}_{ \Xi^{++}_{cc}}\bar u_{ \Xi^{++}_{cc}}(p,s).
\end{eqnarray}
 These matrix elements are used in Eq.(\ref{PhysSide}). After applying the summations over Dirac spinors, 
\begin{eqnarray}\label{Spinors}
\sum_{s'} u_{\Xi^{+}_{c}} (p',s')~\bar{u}_{\Xi^{+}_{c}}
(p',s')&=&\slashed{p}~'+m_{\Xi^{+}_{c}},\notag \\
\sum_{s} u_{ \Xi^{++}_{cc}}(p,s)~\bar{u}_{ \Xi^{++}_{cc}}(p,s)&=&\slashed
p+m_{ \Xi^{++}_{cc}},
\end{eqnarray}
we obtain a representation for the  hadronic side. Finally, we use the double Borel transformation \cite{Aliev:2006gk}:

\begin{eqnarray}\label{BorelQCD2}
	\mathbf{\widehat{B}}\frac{1}{(p^{2}-s)^m} \frac{1}{(p'^{2}-s')^n}\longrightarrow (-1)^{m+n}\frac{1}{\Gamma[m]\Gamma[n]} \frac{1}{(M^2)^{m-1}}\frac{1}{(M'^2)^{n-1}}e^{-s/M^2} e^{-s'/M'^2},
\end{eqnarray}
where $ M^2 $ and $M'^{2}$ are Borel parameters that will be fixed in numerical analysis section. As previously mentioned, the Borel transformation effectively removes the influence of higher resonances and continuum, while amplifying the contributions of ground states in the initial and final channels.  After applying the double Borel transformation, we have
\begin{eqnarray}\label{Physical Side structures}
&&\mathbf{\widehat{B}}~\Pi_{\mu}^{\mathrm{Phys.}}(p,p',q)=\lambda_{ \Xi^{++}_{cc}}\lambda_{\Xi^{+}_{c}}~e^{-\frac{m_{ \Xi^{++}_{cc}}^2}{M^2}}
~e^{-\frac{m_{\Xi^{+}_{c}}^2}{M'^{2}}}\Bigg[F_{1}\bigg(m_{ \Xi^{++}_{cc}} m_{\Xi^{+}_{c}} \gamma_{\mu}+m_{ \Xi^{++}_{cc}} \slashed{p}' \gamma_{\mu}+m_{\Xi^{+}_{c}}\gamma_{\mu}\slashed {p}+\slashed {p}'\gamma_\mu\slashed {p}\bigg)+\notag\\
&&F_2\bigg(\frac{m_{\Xi^{+}_{c}}}{m_{ \Xi^{++}_{cc}}}p_\mu\slashed {p}+\frac{1}{m_{ \Xi^{++}_{cc}}}p_{\mu}\slashed {p}' \slashed {p}+m_{\Xi^{+}_{c}}p_\mu +p_\mu\slashed {p}'\bigg)+ F_3\bigg(\frac{1}{m_{\Xi^{+}_{c}}} p'_{\mu} \slashed {p}' \slashed{p}+p'_\mu\slashed {p}'+p'_\mu\slashed {p}+m_{ \Xi^{++}_{cc}}p'_\mu\bigg)-\notag\\
&& G_1\bigg(m_{ \Xi^{++}_{cc}} m_{\Xi^{+}_{c}} \gamma_{\mu}\gamma_{5}+m_{ \Xi^{++}_{cc}}\slashed {p}'\gamma_\mu\gamma_5-m_{\Xi^{+}_{c}}\gamma_\mu\slashed {p}\gamma_5-\slashed {p}'\gamma_\mu\slashed {p}\gamma_5\bigg)- G_2\bigg(p_\mu\slashed {p}'\gamma_5+m_{\Xi^{+}_{c}}p_\mu\gamma_5-\frac{m_{\Xi^{+}_{c}}}{m_{ \Xi^{++}_{cc}}}p_\mu\slashed {p}\gamma_5-\frac{1}{m_{ \Xi^{++}_{cc}}} p_{\mu} \slashed {p}' \slashed
{p}\gamma_{5}\bigg)\notag\\
&&-G_3\bigg(\frac{m_{ \Xi^{++}_{cc}}}{m_{\Xi^{+}_{c}}}p'_\mu\slashed {p}'\gamma_5+m_{ \Xi^{++}_{cc}}p'_\mu\gamma_5-\frac{1}{m_{\Xi^{+}_{c}}} p'_{\mu}
\slashed {p}'\slashed{p}\gamma_{5}-p'_\mu\slashed {p}\gamma_5\bigg)\Bigg]+\cdots~.
\end{eqnarray}
\\
\subsection{QCD side}
To obtain the QCD side, it is necessary to calculate the correlation function in Eq. (\ref{CorFunc}) by incorporating the interpolating currents of the initial and final baryons.  We use the interpolating currents for the singly and doubly heavy baryons with spin-parity $J^P=(\frac{1}{2})^+$ as follows \cite{Aliev:2010yx}:
\begin{eqnarray} \label{Current}
	{\cal J}^{\Xi^{+}_{c}}(x)&&=\frac{1}{\sqrt{6}}~\epsilon_{abc}\Bigg\{2 \Big(u^{aT}(x)Cs^{b}(x)\Big)\gamma_{5}c^{c}(x) + \Big(u^{aT}(x)Cc^{b}(x)\Big)\gamma_{5}s^{c}(x) +\Big(c^{aT}(x)Cs^{b}(x)\Big)\gamma_{5}u^{c}(x) \notag \\
	&&+
	2 \beta\Big(u^{aT}(x)C\gamma_{5}s^{b}(x)\Big)c^{c}(x)  + \beta\Big(u^{aT}(x)C\gamma_{5}c^{b}(x)\Big)s^{c}(x)    + \beta\Big(c^{aT}(x)C\gamma_{5}s^{b}(x)\Big)u^{c}(x) \Bigg\},~
\end{eqnarray}

and
\begin{eqnarray} \label{Currents}
	{\cal J}^{\Xi^{++}_{cc}}(x)&&= \sqrt{2}~\epsilon_{abc}\Bigg\{ \Big(c^{aT}(x)Cu^{b}(x)\Big)\gamma_{5}c^{c}(x) + \beta\Big(c^{aT}(x)C\gamma_{5}u^{b}(x)\Big)c^{c}(x) \Bigg\},~
\end{eqnarray}
where the superindices  $a,~b$ and $c$  are color indices, $C$ is the charge
conjugation operator and $c$, $u$ and $s$ represent the  charm, up and strange quark fields. The parameter $\beta$  is an arbitrary mixing parameter to be fixed from the numerical analysis, with $\beta=-1$ corresponding to the Ioffe current.

We start the calculation of the QCD side by substituting the interpolating currents of the initial, ${\cal J}^{\Xi^{++}_{cc}}$, and final, ${\cal J}^{\Xi^{+}_{c}}$, baryons along with the transition current \(J^{tr}_\mu\) into the correlation function. By applying Wick’s theorem to perform all possible contractions of the quark fields, we express the correlation function in relation to the propagators for both the heavy and light quarks. The final result is:

\begin{eqnarray} \label{ term}
	&&\Pi^{OPE}_{\mu}=i^2 \int d^4x e^{-ipx}\int d^4y e^{ip'y} \frac{1}{\sqrt{3}} \epsilon_{a'b'c'} \epsilon_{abc}\Bigg\{2 \gamma_5~ S^{c a'}_c(y-x) S'^{a b'}_u(y-x) S^{b i}_s(y) \gamma_\mu (1-\gamma_5) S^{i c'}_c(-x) \gamma_5\notag\\
	&&+2 Tr[S^{a b'}_u(y-x) S'^{i a'}_c(-x) (1-\gamma_5)\gamma_\mu S'^{b i}_s(y)] ~\gamma_5 S^{c c'}_c(y-x)\gamma_5-2
	\beta Tr[S^{b i}_s(y)  \gamma_\mu (1-\gamma_5) S^{i a'}_c(-x)~\gamma_5 S'^{a b'}_u(y-x)] \notag\\
	&&\gamma_5 S^{c c'}_c(y-x)+2 \beta \gamma_5~ S^{c a'}_c(y-x) \gamma_5 S'^{a b'}_u(y-x) S^{b i}_s(y) \gamma_\mu (1-\gamma_5) S^{i c'}_c(-x) - Tr[S'^{b a'}_c(y-x) S^{a b'}_u(y-x)] \gamma_5 S^{c i}_s(y) \notag\\
	&&\gamma_\mu(1-\gamma_5) S^{i c'}_c(-x) ~\gamma_5+\gamma_5~ S^{c i}_s(y) \gamma_\mu (1-\gamma_5) S^{i a'}_c(-x) \gamma_5 S'^{a b'}_u(y-x)  S^{b c'}_c(y-x) -\beta \gamma_5 S^{c i}_s(y) \gamma_\mu (1-\gamma_5) S^{i c'}_c(-x)  \notag\\
	&&Tr[  \gamma_5 S'^{b a'}_c(y-x) S^{a b'}_u(y-x) ]+\beta \gamma_5 S^{c i}_s(y)  \gamma_\mu(1-\gamma_5)   S^{i a'}_c(-x) \gamma_5 S'^{a b'}_u(y-x)~S^{b c'}_c(y-x) + \gamma_5 S^{c b'}_u(y-x) S'^{a a'}_c(y-x)  \notag\\
	&&S^{b i}_s(y) \gamma_\mu (1-\gamma_5) S'^{i c'}_c(-x)  \gamma_5  + \gamma_5 S^{c b'}_u(y-x) S'^{i a'}_c(-x)   (1-\gamma_5)\gamma_\mu S'^{b i}_s(y)  S^{a c'}_c(y-x)  \gamma_5 +\beta \gamma_5 S^{c b'}_u(y-x)  \gamma_5  S'^{a a'}_c(y-x) \notag\\
	&&S^{b i}_s(y)  \gamma_\mu(1-\gamma_5)  ~S^{i c'}_c(-x) +\beta \gamma_5 S^{c b'}_u(y-x)  \gamma_5  S'^{i a'}_c(-x)  (1-\gamma_5) \gamma_\mu S'^{b i}_s(y)  ~S^{a c'}_c(y-x) -2 \beta Tr[S'^{a b'}_u(y-x)  \gamma_5  S^{b i}_s(y)  \gamma_\mu\notag\\
	&& (1-\gamma_5) S^{i a'}_c(-x) ] S^{c c'}_c(y-x)  \gamma_5  + 2 \beta Tr[S'^{b a'}_c(y-x)  \gamma_5  S^{a b'}_u(y-x) ] S^{b i}_s(y)  \gamma_\mu (1-\gamma_5)   ~S^{i c'}_c(-x)  \gamma_5 +2 \beta^2 S^{c a'}_c(y-x)  \gamma_5    \notag\\
	&&S'^{a b'}_u(y-x)  \gamma_5 S^{b i}_s(y)   \gamma_\mu (1-\gamma_5) S^{i c'}_c(-x)  - 2 \beta^2 Tr[S'^{a b'}_u(y-x)  \gamma_5  S'^{i a'}_c(-x)  (1-\gamma_5)  \gamma_\mu   S'^{b i}_s(y)   \gamma_5] S^{c c'}_c(y-x)  \notag\\
	&&- 2 \beta  Tr[S'^{a b'}_u(y-x)  \gamma_5  S^{b a'}_c(y-x)] S^{c i}_s(y) \gamma_\mu   (1-\gamma_5)  S^{i c'}_c(-x)     \gamma_5  -  \beta S^{c i}_s(y)  \gamma_\mu (1-\gamma_5) S^{i a'}_c(-x)   S'^{a b'}_u(y)  \gamma_5   S^{b c'}_c(y-x)    \gamma_5] \notag\\
	&&-  \beta^2  Tr[S^{a b'}_u(y-x)  \gamma_5  S'^{b a'}_c(y-x) \gamma_5  ] S^{c i}_s(y) \gamma_\mu   (1-\gamma_5)  S^{i c'}_c(-x)  + \beta^2 S^{c i}_s(y)  \gamma_\mu (1-\gamma_5) S^{i a'}_c(-x)   \gamma_5 S'^{a b'}_u(y-x)    \gamma_5   \notag\\
	&& S^{b c'}_c(y-x)+  \beta  S^{c b'}_u(y-x)   S'^{a a'}_c(y-x) \gamma_5  S^{b i}_s(y) \gamma_\mu   (1-\gamma_5)  S^{i c'}_c(-x)  \gamma_5  + \beta  S^{c b'}_u(y-x) S'^{i a'}_c(-x)   (1-\gamma_5) \gamma_\mu   S'^{b i}_s(y)    \gamma_5   \notag\\
	&& S^{a c'}_c(y-x)  \gamma_5  +  \beta^2  S^{c b'}_u(y-x)  \gamma_5 S'^{i a'}_c(-x)    (1-\gamma_5)  \gamma_\mu  S'^{b i}_s(y)  \gamma_5   S^{a c'}_c(y-x)  + \beta^2 S^{c b'}_u(y-x) \gamma_5  S'^{a a'}_c(y-x)  \gamma_5 S^{b i}_s(y)   \notag\\
	&& \gamma_\mu (1-\gamma_5)  S^{i c'}_c(-x) \Bigg\},
\end{eqnarray}

where $S'=C S^T C$.  For the light and heavy quark propagators, we utilize \cite{Agaev:2020zad}:

\begin{eqnarray}\label{LightProp}
S_{q}^{ab}(x)&=&i\delta _{ab}\frac{\slashed x}{2\pi ^{2}x^{4}}-\delta _{ab}%
\frac{m_{q}}{4\pi ^{2}x^{2}}-\delta _{ab}\frac{\langle\overline{q}q\rangle}{12} +i\delta _{ab}\frac{\slashed xm_{q}\langle \overline{q}q\rangle }{48}%
-\delta _{ab}\frac{x^{2}}{192}\langle \overline{q}g_{}\sigma
Gq\rangle+
i\delta _{ab}\frac{x^{2}\slashed xm_{q}}{1152}\langle \overline{q}g_{}\sigma Gq\rangle \notag\\
&-&i\frac{g_{}G_{ab}^{\alpha \beta }}{32\pi ^{2}x^{2}}\left[ \slashed x{\sigma _{\alpha \beta }+\sigma _{\alpha \beta }}\slashed x\right]-i\delta _{ab}\frac{x^{2}\slashed xg_{}^{2}\langle
\overline{q}q\rangle ^{2}}{7776} -\delta _{ab}\frac{x^{4}\langle \overline{q}q\rangle \langle
g_{}^{2}G^{2}\rangle }{27648}+\ldots,
\end{eqnarray}
and
\begin{eqnarray}\label{HeavyProp}
&&S_{Q}^{ab}(x)=i\int \frac{d^{4}k}{(2\pi )^{4}}e^{-ikx}\Bigg
\{\frac{\delta_{ab}\left( {\slashed k}+m_{Q}\right) }{k^{2}-m_{Q}^{2}}-\frac{g_{}G_{ab}^{\mu \nu}}{4}\frac{\sigma _{\mu\nu }\left( {%
\slashed k}+m_{Q}\right) +\left( {\slashed k}+m_{Q}\right) \sigma
_{\mu\nu}}{(k^{2}-m_{Q}^{2})^{2}} +\frac{g_{}^{2}G^{2}}{12}\delta _{ab}m_{Q}\frac{k^{2}+m_{Q}{\slashed k}}{%
(k^{2}-m_{Q}^{2})^{4}}\notag\\
&&+\ldots\Bigg \},
\end{eqnarray}

where $G_{\mu\nu}$ is the gluon field strength tensor,  $G_{ab}^{\mu \nu }=G_{A}^{\mu\nu
}t_{ab}^{A}$, $t^A=\lambda^A/2$ and $G^{2}=G_{A}^{\mu\nu} G_{\mu \nu }^{A}$. $\lambda
^{A}$ are the Gell-Mann matrices and A takes  amounts from 1 to 8. In the QCD side, the correlation function is calculated in the deep Euclidean region by utilizing the operator product expansion (OPE). 
In Wilson's OPE, each term in the light and heavy quark propagators introduces an operator with a specific mass dimension. The bare-loop term with \(d=0\) represents a perturbative contribution, while the corrections arise from operators of various dimensions: \(d=3\) like \(\langle \overline{q}q\rangle\), \(d=4\) such as \(\langle G^2\rangle\), and \(d=5\) like \(\langle \overline{q}g_{}\sigma Gq\rangle\),  are considered as nonperturbative terms. By computing the correlation function, we obtain the consequence containing both perturbative and nonperturbative corrections across various mass dimensions. Our calculations incorporate nonperturbative operators up to five mass dimensions. In this study, the masses of light quarks are assumed to be zero. After evaluating the integrals, we find the correlation function in terms of 24 different Lorentz structures as shown:
\begin{eqnarray}\label{Structures}
&&\Pi_{\mu}^{\mathrm{OPE}}(p,p',q)=\Pi^{\mathrm{OPE}}_{\slashed{p}' \gamma_{\mu}\slashed{p}}(p^{2},p'^{2},q^{2})~\slashed{p}' \gamma_{\mu}\slashed{p}+
\Pi^{\mathrm{OPE}}_{p_{\mu} \slashed {p}'\slashed {p}}(p^{2},p'^{2},q^{2})~p_{\mu} \slashed {p}'\slashed {p}+
\Pi^{\mathrm{OPE}}_{p_{\mu}' \slashed {p}'\slashed {p}}(p^{2},p'^{2},q^{2})~p_{\mu}' \slashed {p}'\slashed {p}+\Pi^{\mathrm{OPE}}_{p'_\mu\slashed {p}'\gamma_5}(p^{2},p'^{2},q^{2})\notag\\
&&p'_\mu\slashed {p}'\gamma_5+
\Pi^{\mathrm{OPE}}_{p'_\mu\slashed {p}'\slashed{p}\gamma_5}(p^{2},p'^{2},q^{2})~p'_\mu\slashed {p}'\slashed{p}\gamma_5+
\Pi^{\mathrm{OPE}}_{\slashed {p}'\gamma_\mu\gamma_5}(p^{2},p'^{2},q^{2})~\slashed {p}'\gamma_\mu\gamma_5+
\Pi^{\mathrm{OPE}}_{\slashed {p}'\gamma_\mu\slashed {p}\gamma_5}(p^{2},p'^{2},q^{2})~\slashed {p}'\gamma_\mu\slashed {p}\gamma_5+\Pi^{\mathrm{OPE}}_{p_{\mu} \slashed {p}' \slashed{p}\gamma_{5}}(p^{2},p'^{2},q^{2})\notag\\
&&p_{\mu} \slashed {p}' \slashed{p}\gamma_{5}+
 \Pi^{\mathrm{OPE}}_{\slashed{p}' \gamma_{\mu}}(p^{2},p'^{2},q^{2})~\slashed{p}' \gamma_{\mu}+
\Pi^{\mathrm{OPE}}_{p_\mu\slashed {p}'\gamma_5}(p^{2},p'^{2},q^{2})~p_\mu\slashed {p}'\gamma_5+
\Pi^{\mathrm{OPE}}_{p'_\mu\slashed {p}'}(p^{2},p'^{2},q^{2})~p'_\mu\slashed {p}'+
\Pi^{\mathrm{OPE}}_{p_\mu\slashed {p}'}(p^{2},p'^{2},q^{2})~p_\mu\slashed {p}'+\notag\\
&&\Pi^{\mathrm{OPE}}_{\gamma_\mu\slashed {p}\gamma_5}(p^{2},p'^{2},q^{2})~\gamma_\mu\slashed {p}\gamma_5+
\Pi^{\mathrm{OPE}}_{\gamma_{\mu}}(p^{2},p'^{2},q^{2})~\gamma_{\mu}+
\Pi^{\mathrm{OPE}}_{\gamma_{\mu}\slashed {p}}(p^{2},p'^{2},q^{2})~\gamma_{\mu}\slashed {p}+
\Pi^{\mathrm{OPE}}_{ \gamma_{\mu}\gamma_{5}}(p^{2},p'^{2},q^{2}) ~\gamma_{\mu}\gamma_{5}+\Pi^{\mathrm{OPE}}_{p_\mu\slashed {p}\gamma_5}(p^{2},p'^{2},q^{2})\notag\\
&&p_\mu\slashed {p}\gamma_5+
\Pi^{\mathrm{OPE}}_{p'_\mu\slashed {p}\gamma_5}(p^{2},p'^{2},q^{2})~p'_\mu\slashed {p}\gamma_5+
\Pi^{\mathrm{OPE}}_{p'_\mu\slashed {p}}(p^{2},p'^{2},q^{2})~p'_\mu\slashed {p}+
\Pi^{\mathrm{OPE}}_{p_\mu\slashed {p}}(p^{2},p'^{2},q^{2})~p_\mu\slashed {p}+\Pi^{\mathrm{OPE}}_{p'_\mu}(p^{2},p'^{2},q^{2})~p'_\mu+\notag\\
&&
\Pi^{\mathrm{OPE}}_{p'_\mu\gamma_5}(p^{2},p'^{2},q^{2})~p'_\mu\gamma_5+
\Pi^{\mathrm{OPE}}_{p_\mu}(p^{2},p'^{2},q^{2})~p_\mu+
\Pi^{\mathrm{OPE}}_{p_\mu\gamma_5}(p^{2},p'^{2},q^{2})~p_\mu\gamma_5.
\end{eqnarray}
The invariant functions $\Pi^{\mathrm{OPE}}_i(p^{2},p'^{2},q^{2})$ (where $i$ represents distinct structures) are defined in terms of double dispersion integrals:
\begin{eqnarray}\label{PiQCD}
\Pi^{\mathrm{OPE}}_i(p^{2},p'^{2},q^{2})&=&\int_{s_{min}}^{\infty}ds
\int_{s'_{min}}^{\infty}ds'~\frac{\rho
^{\mathrm{OPE}}_i(s,s',q^{2})}{(s-p^{2})(s'-p'^{2})} ,
\end{eqnarray}
where $s_{min}=(m_c+m_c)^{2}$, $s'_{min}=(m_c)^{2}$ and $\rho_i^{\mathrm{OPE}}(s,s',q^{2})$ explain the spectral densities, achieved by $\rho_i^{\mathrm{OPE}}(s,s',q^{2})=\frac{1}{\pi}Im\Pi^{OPE}_i(p^2,p'^2,q^2)$. 
After applying the quark hadron duality assumption, the integrals' upper limits will be modified to $s_0$ and $s'_0$, representing the continuum thresholds of the initial and final baryon states. We can express the spectral densities as follows:
\begin{equation} \label{Rhoqcd}
\rho^{\mathrm{OPE}}_i(s,s',q^{2})=\rho_i ^{Pert.}(s,s',q^{2})+\sum_{n=3}^{5}\rho_{i}^{n}(s,s',q^{2}),
\end{equation} 
where $\rho_i ^{Pert.}(s,s',q^{2})$ denotes the perturbative part of calculation. We define $\sum_{n=3}^{5}\rho_{i}^{n}(s,s',q^{2})$ for all mass dimensions of nonperturbative part of calculation including quark condensate, gluon condensate and quark-gluon mixed condensate, respectively. 
 Additionally, we apply the double Borel transformation to the QCD side and perform continuum subtraction based on the quark hadron duality assumption. Consequently, we obtain:
\begin{eqnarray}\label{qcd part2}
&&\Pi^{\mathrm{OPE}}_i (M^2,M'^2,s_0,s'_0,q^2)=\int _{s_{min}}^{s_0} ds\int _{s'_{min}}^{s'_0}ds' e^{-s/M^2} e^{-s'/M'^2}\rho
^{\mathrm{OPE}}_{i}(s,s',q^{2}),\nonumber\\
\end{eqnarray}
In the Appendix, we provide the components of $\rho_{i}(s,s^{\prime},q^2)$ as an example for the $\gamma_{\mu} \gamma_5$ structure.

We then derive the necessary sum rules for the form factors to be used in numerical calculations by matching the corresponding coefficients of the various Lorentz structures from the hadronic and QCD sides. The sum rules for the desired form factors are  expressed in terms of the baryon masses and residues,  the QCD parameters such as the strong coupling constant, quark masses, quark condensate, gluon condensates, ..., and the auxiliary parameters $M^2$, $M'^2$, $s_0$, $s'_0$ and $\beta$.

\section{Numerical Analysis of the form factors}\label{Sec3}

The form factors contain all the necessary information to determine the decay width of the weak transition under investigation. The main aim in this respect is to find the $ q^2 $ dependence of the form factors in the whole physical region in this section. The input parameters required for the numerical  calculations are listed in Table\ \ref{inputParameter}.

\begin{table}[h!]
\caption{The input parameters used in our numerical calculation.}\label{inputParameter}
\begin{tabular}{|c|c|}
\hline 
Parameters                                             &  Values  \\
\hline 
$ m_c$                                                 & $(1.27\pm0.02)~ \mathrm{GeV}$ \cite{ParticleDataGroup:2020ssz}\\
$ m_e $                                                & $ 0.51~\mathrm{MeV}$ \cite{ParticleDataGroup:2020ssz}\\
$ m_\mu $                                              & $105~\mathrm{MeV}$ \cite{ParticleDataGroup:2020ssz}\\
$ m_{ \Xi^{++}_{cc}}$                                       & $ (3.62\pm0.0015)~ \mathrm{GeV}$ \cite{ParticleDataGroup:2020ssz}\\
$ m_{\Xi^{+}_{c}} $                                      & $ (2.46\pm0.00023)~\mathrm{GeV}$  \cite{ParticleDataGroup:2020ssz} \\
$ G_{F} $                                              & $ 1.17\times 10^{-5} ~\mathrm{GeV^{-2}}$ \cite{ParticleDataGroup:2020ssz}\\
$ |V_{sc}| $                                             & $ (0.974\pm 0.006 )$  \cite{ParticleDataGroup:2020ssz}\\
$ m^2_0 $                                              & $ (0.8\pm0.2)~ \mathrm{GeV^2}$\cite{Belyaev:1982sa,Belyaev:1982cd,Ioffe:2005ym} \\
$\tau_{ \Xi^{++}_{cc}} $                                    & $ 2.56\pm 0.27\times 10^{-13}~ s$ \cite{ParticleDataGroup:2020ssz}\\

$\langle \bar{u} u\rangle$         & $-(0.24\pm0.01)^3 ~\mathrm{GeV^3}$ \cite{Belyaev:1982sa,Belyaev:1982cd} \\
$\langle \bar{s} s\rangle$           & $(0.8\pm0.1) \langle \bar{u}u\rangle~ \mathrm{GeV^3}$ \cite{Belyaev:1982sa,Belyaev:1982cd} \\

$\langle0|\frac{1}{\pi}\alpha_s G^2|0\rangle$          &$ (0.012\pm0.004)~\mathrm{GeV^4}$ \cite{Belyaev:1982sa,Belyaev:1982cd,Ioffe:2005ym}\\
$\lambda_{ \Xi^{++}_{cc}}$                               &$0.16\pm 0.04 ~ \mathrm{GeV^3}$ \cite{ShekariTousi:2024mso}\\
$\lambda_{\Xi^{+}_{c}}$                               & $0.027\pm 0.008~  \mathrm{GeV^3}$ \cite{Wang:2010fq}\\
\hline
\end{tabular}
\end{table}
The sum rules for the form factors include five additional auxiliary parameters known as the Borel parameters $M^2$ and $M'^2$, as well as the continuum thresholds $s_0$ and $s'_0$ and the mixing parameter $\beta$. According to the standards of the method, the form factors as physical quantities are expected to be possibly insensitive to these parameters. In practice, however, there appear residual dependencies on these helping parameters. As a result, specific ranges are chosen for these parameters to ensure that the form factors  depend relatively weakly on them.  These windows are obtained by  taking into account the standard  requirements. These criteria include weak dependence of the results on auxiliary parameters,
pole dominance and convergence of the OPE. The upper bound for the Borel mass parameters $M^2$ and $M'^2$, are set by ensuring that the pole contribution is larger than the contributions coming from the higher states and  continuum. To this end, we demand
\begin{equation} \label{PC}
PC=\frac{\Pi^{OPE}(M^2,M'^2,s_0,s'_0)}{\Pi^{OPE}(M^2,M'^2,{\infty},{\infty})}\geq 0.5.
\end{equation}
 The lower bounds of the Borel mass parameters $M^2$ and $M'^2$  are determined by the condition that the OPE series must be convergent. We require that the perturbative contribution be greater than the nonperturbative contribution, and that the contributions of nonperturbative operators decrease with increasing dimension. To ensure this, we impose the following condition:

\begin{equation} \label{PC2}
R(M^2, M'^2)=\frac{\Pi^{{OPE}-dim5}(M^2,M'^2,s_0,s'_0)}{\Pi^{OPE}(M^2,M'^2,s_0,s'_0)}\leq0.05.
\end{equation}
With these requirements, the working regions for the Borel parameters are determined as $4~\mathrm{GeV^2}\leq M^2 \leq 6~\mathrm{GeV^2}$ and $3~\mathrm{GeV^2} \leq M'^2 \leq 5~\mathrm{GeV^2}$.
The continuum thresholds $s_{0}$ and $s'_0$ are not arbitrary, and their values are carefully selected to ensure that the integrals exclude any contributions from the excited states in the calculations. These threshold parameters are also determined by evaluating the stability of the sum rules within the chosen intervals of the Borel mass parameters $M^ 2$  and $M'^2$. To achieve the best stability and physical consistency, therefore, $s_0 $ and $s'_0$  are set based on the requirement that they accurately reflect the onset of the continuum in the spectrum, ensuring that the contributions from higher states and continuum are appropriately suppressed. This careful selection leads to the working regions for the continuum thresholds, which are determined as follows:
\begin{eqnarray}
&&(m_{ \Xi^{++}_{cc}}+0.25)^2~ \mathrm{GeV^2} \leq s_{0} \leq (m_{ \Xi^{++}_{cc}}+0.62)^2~ \mathrm{GeV^2},\notag\\
\mbox{and} \notag\\
&&(m_{\Xi^{+}_{c}}+0.3)^2~\mathrm{GeV^2}\leq s'_{0} \leq (m_{\Xi^{+}_{c}}+0.5)^2~ \mathrm{GeV^2},
\end{eqnarray}

\begin{figure}[h!] 
	\includegraphics[totalheight=5cm,width=5.8cm]{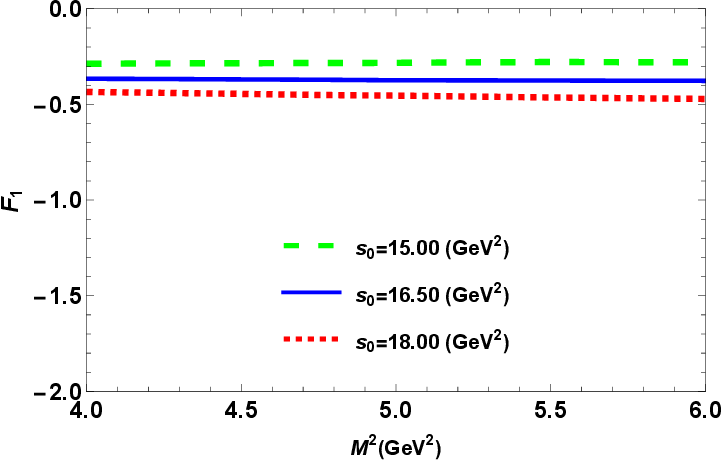}
	\includegraphics[totalheight=5cm,width=5.8cm]{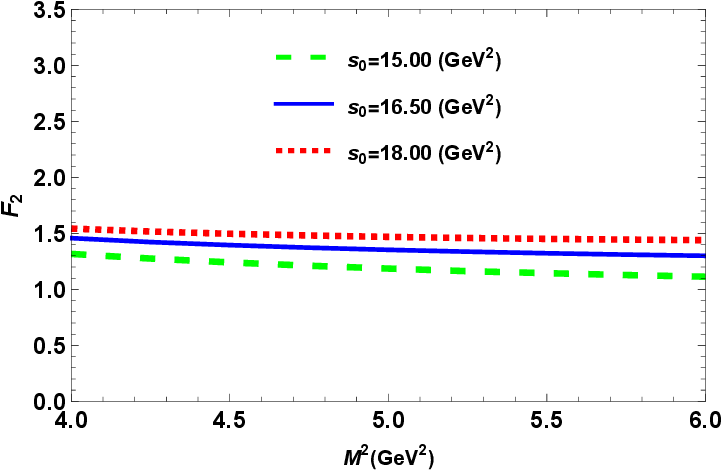}
	\includegraphics[totalheight=5cm,width=5.8cm]{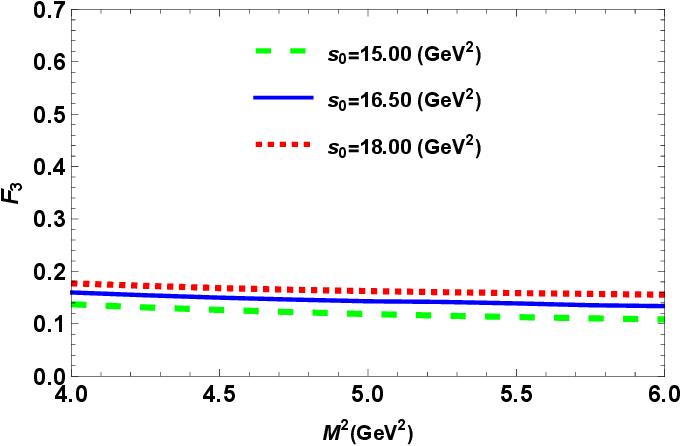}
	\includegraphics[totalheight=5cm,width=5.8cm]{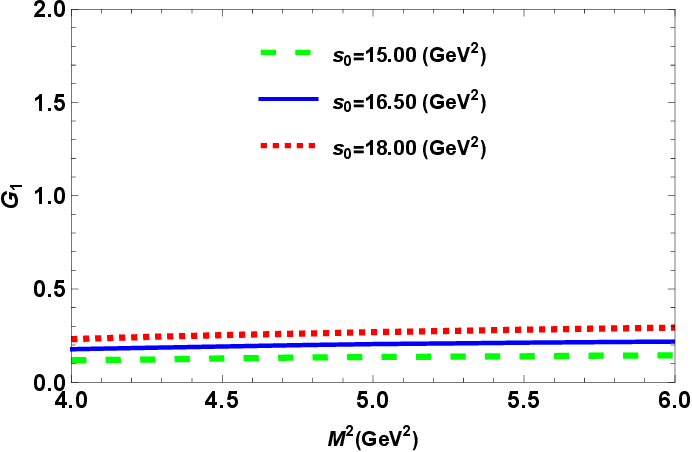}
	\includegraphics[totalheight=5cm,width=5.8cm]{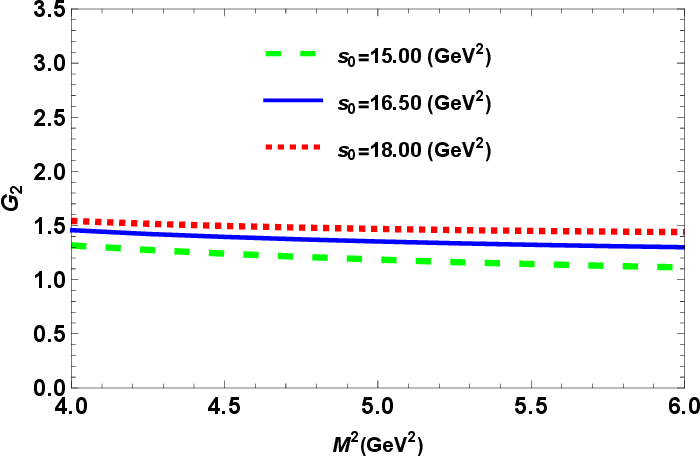}
	\includegraphics[totalheight=5cm,width=5.8cm]{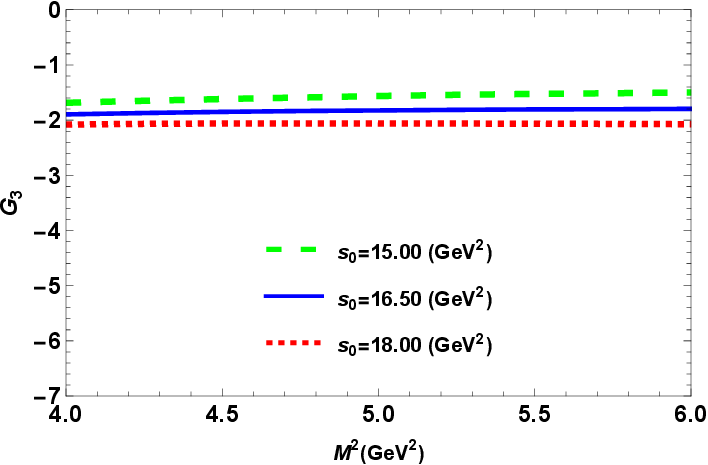}
	\caption{Variations of the form factors  as  functions of the Borel parameter $M^2$, at the different values of the parameter $s_0$, $q^2=0$ and at the central  values of other auxiliary parameters. The graphs correspond to the  structures $\slashed {p}'\gamma_{\mu}  $, $p_{\mu}\slashed {p'}\slashed {p}$, $p'_{\mu}\slashed {p'}\slashed {p}$, $\gamma_{\mu} \gamma_5$, $p_{\mu}\slashed {p}'\slashed {p}\gamma_5$ and $p'_{\mu}\gamma_5$ for $F_1$, $F_2$, $F_3$, $G_1$, $G_2$ and $G_3$, respectively.}\label{Fig:BorelM}
\end{figure}

\begin{figure}[h!]
	\includegraphics[totalheight=5cm,width=5.8cm]{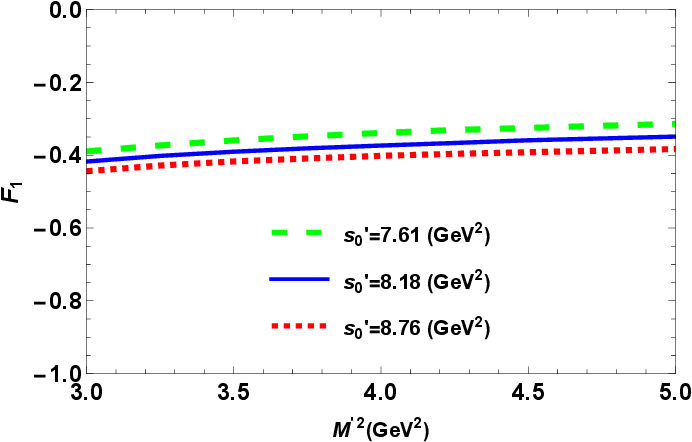}
	\includegraphics[totalheight=5cm,width=5.8cm]{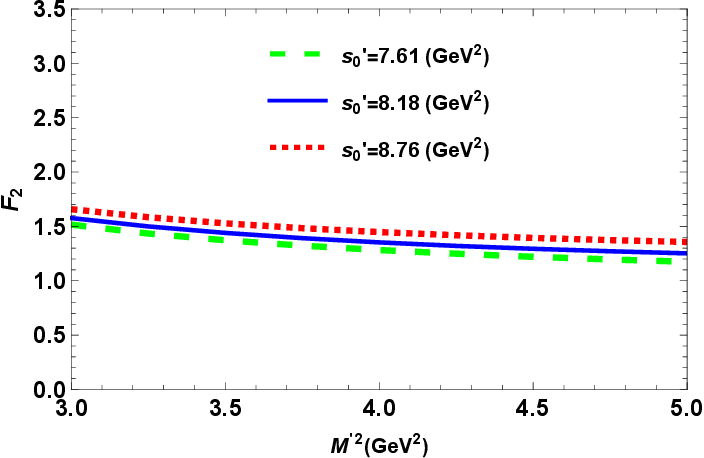}
	\includegraphics[totalheight=5cm,width=5.8cm]{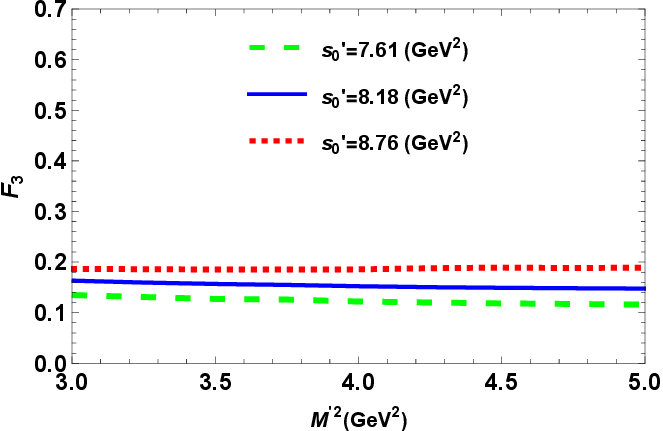}
	\includegraphics[totalheight=5cm,width=5.8cm]{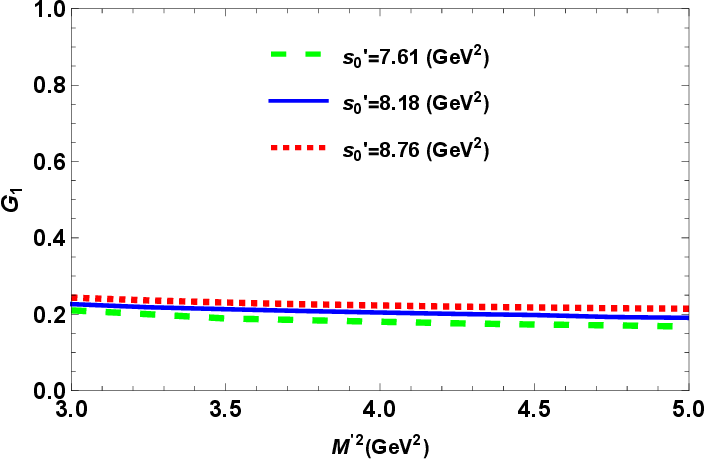}
	\includegraphics[totalheight=5cm,width=5.8cm]{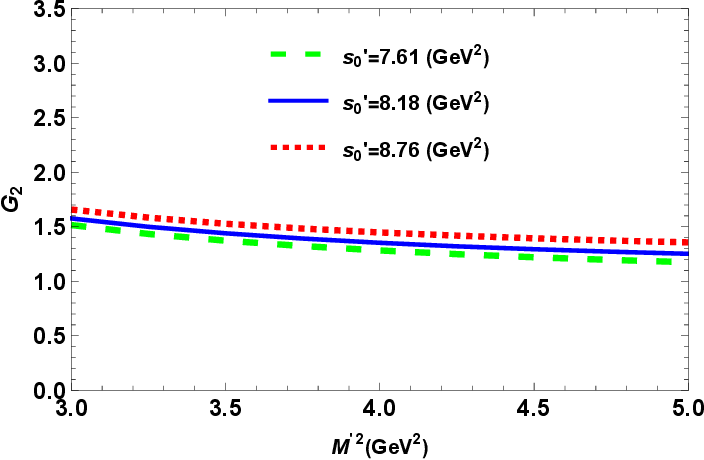}
	\includegraphics[totalheight=5cm,width=5.8cm]{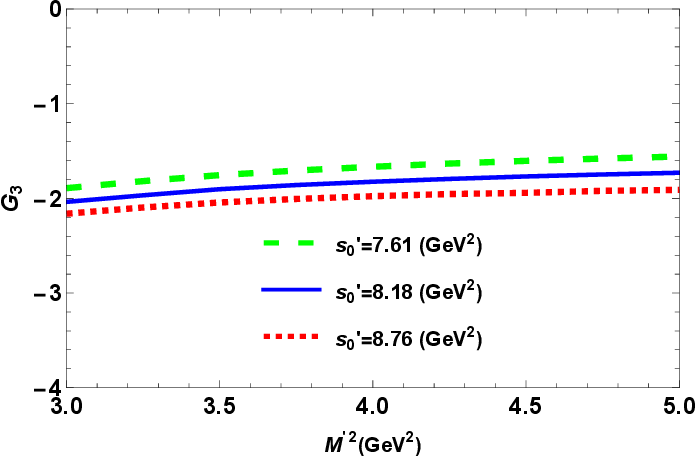}
	\caption{Variations of the form factors  as  functions of the Borel parameter $M'^2$, at the different values of the parameter $s'_0$, $q^2=0$ and at the central values of other auxiliary parameters. The graphs correspond to the  structures $\slashed {p}'\gamma_{\mu}  $, $p_{\mu}\slashed {p'}\slashed {p}$, $p'_{\mu}\slashed {p'}\slashed {p}$, $\gamma_{\mu} \gamma_5$, $p_{\mu}\slashed {p}'\slashed {p}\gamma_5$ and $p'_{\mu}\gamma_5$ for $F_1$, $F_2$, $F_3$, $G_1$, $G_2$ and $G_3$, respectively.} \label{Fig:BorelMM}
\end{figure}

As shown in Figs. \ref{Fig:BorelM} and \ref{Fig:BorelMM}, the form factors exhibit notable stability when considering variations of $M^2$, $M'^2$, $s_0$, and $s'_0$ within their respective intervals. This stability indicates that the chosen parameters are well-optimized and that the results are reliable across the defined ranges.

\begin{figure}[h!]
	\includegraphics[totalheight=6cm,width=8cm]{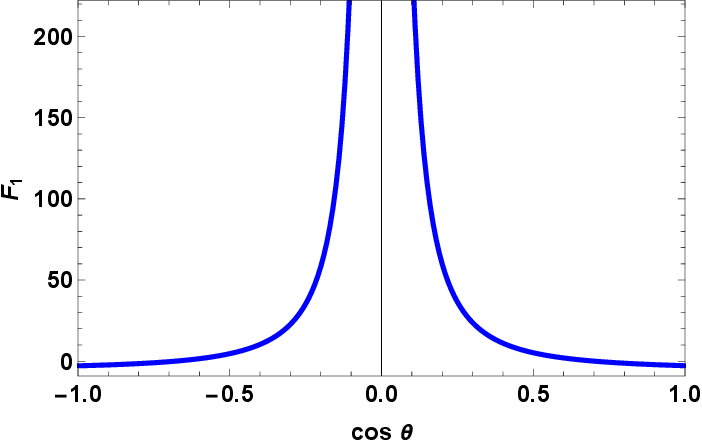}
	\includegraphics[totalheight=6cm,width=8cm]{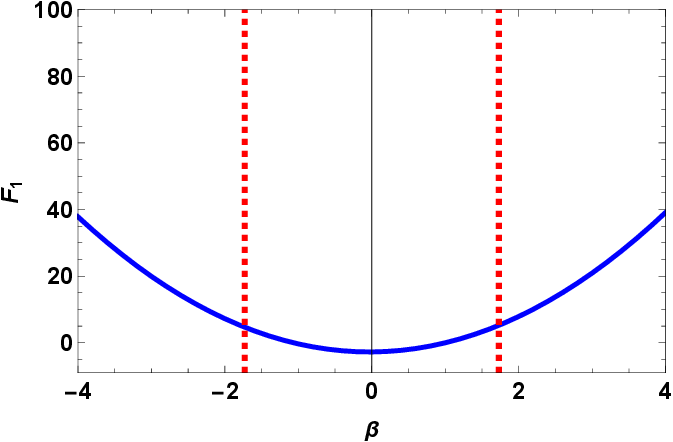}
	\caption{The variations of $F_1$ form factor as  functions of $\cos \theta$ (or $x$) and $\beta$ at the central  values of other auxiliary parameters and at $q^2=0$, corresponding to the $\slashed {p}'\gamma_{\mu}  $ structure.} \label{Fig:f1x}
\end{figure}

In addition to the parameters $M^2$, $M'^2$, $s_0$ and $s'_0$, the parameter $\beta$ plays a crucial role, but it does not have a restricted range and can theoretically extend from $-\infty$ to $\infty$. To determine the optimal working region for the $\beta$  parameter, we use the transformation $x = \cos\theta$, where $\theta = \tan^{-1}\beta$. The selected range for $x$ is chosen to maintain the stability of the form factors without causing significant changes. For instance, Fig. \ref{Fig:f1x} illustrates the variations of the $F_1$ form factor as a function of $ \cos\theta$ (or $x$). Based on this illustration, we constrain the $x$ parameter to the intervals $-1\leq x \leq -0.5$ and
$0.5\leq x \leq 1$, which corresponds to $ \beta\in [-1.73, 1.73] $. This constraint is applied to all the six form factors. As shown in Fig. \ref{Fig:f1x}, the $F_1 $ form factor demonstrates minimal variation with $\beta$, particularly  around $ \beta=-1 $, which corresponds to $x = -0.71$  and is associated with the Ioffe current in the identified negative region.

  After fixing the working regions of the auxiliary parameters, we examine the behavior of form factors  with respect to $q^2$. In our study, we find that the form factors are well fitted to the following function:

\begin{equation} \label{fitffunction}
{\cal F}(q^2)=\frac{{\cal
F}(0)}{\displaystyle\left(1-a_1\frac{q^2}{m^2_{ \Xi^{++}_{cc}}}+a_2
\frac{q^4}{m_{ \Xi^{++}_{cc}}^4}+a_3\frac{q^6}{m_{ \Xi^{++}_{cc}}^6}+a_4\frac{q^8}{m_{ \Xi^{++}_{cc}}^8}\right)}.
\end{equation}

The values of the parameters, ${\cal F}(0)$, $a_1$, $a_2$, $a_3$, and $a_4$, acquired utilizing the center of values of the auxiliary parameters, are presented in Table~\ref{Tab:parameterfit}. These values are calculated for the six form factors, $F_1$, $F_2$, $F_3$, $G_1$, $G_2$ and $G_3$, corresponding to structures $\slashed {p}'\gamma_{\mu}  $, $p_{\mu}\slashed {p'}\slashed {p}$, $p'_{\mu}\slashed {p'}\slashed {p}$, $\gamma_{\mu} \gamma_5$, $p_{\mu}\slashed {p}'\slashed {p}\gamma_5$ and $p'_{\mu}\gamma_5$, respectively. It should be noted that the above  parametrization is not unique and different functions and fitting procedures can be applied. One of alternative parametrizations is  to employ the model-independent $z$-series widely discussed in the context of heavy-to-light $B$-meson decays in Refs. \cite{Wang:2015vgv,Cui:2022zwm}.

\begin{table}[h!]
\caption{The fit function parameters for various form factors for $ \Xi^{++}_{cc}\rightarrow \Xi^+_{c} \bar{\ell}\nu_{\ell}$ decay channel.}\label{Tab:parameterfit}
\begin{ruledtabular}
\begin{tabular}{|c|c|c|c|c|c|c|}
            & $F_1(q^2)$ & $F_2(q^2)$  & $F_3(q^2)$   & $G_1(q^2)$ & $G_2(q^2)$  & $G_3(q^2)$       \\
\hline
${\cal F}(q^2=0)$ & $-0.37\pm0.13$        & $1.35\pm0.43$      & $0.16\pm0.06$     & $0.20\pm0.06$  & $1.34\pm0.43$  & $-1.86\pm0.65$  \\
$a_1$           & $1.54$          & $-0.05$            & $1.99$           & $2.79$           & $1.07$            &$ 2.15$              \\
$a_2$           & $-23.84$         & $-111.38$           & $17.17$             & $35.82$          & $-78.83$           & $-27.91$           \\
$a_3$           & $268.50$            & $1329.88$           & $-442.35$          & $-541.92$          & $935.73$           & $449.20$           \\
$a_4$           & $-1174.79$           & $-5206.92$           & $2372.78$            & $2240.25$         & $-3550.64$          & $-2702.99$           \\
\end{tabular}
\end{ruledtabular}
\end{table}
%%%%%%%%%%%%%%
%
\begin{figure}[h!] 
\includegraphics[totalheight=5cm,width=5.8cm]{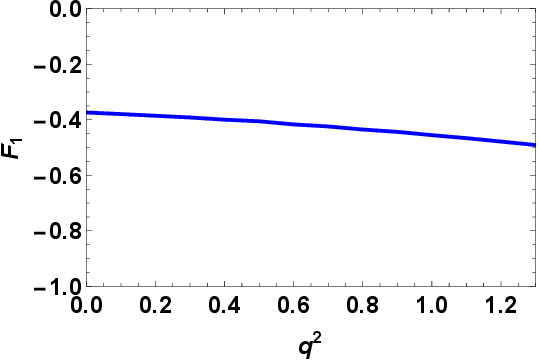}
\includegraphics[totalheight=5cm,width=5.8cm]{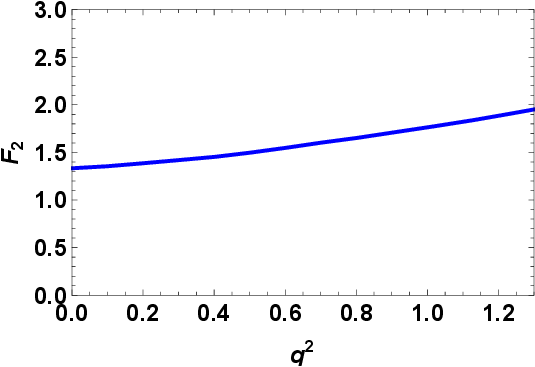}
\includegraphics[totalheight=5cm,width=5.8cm]{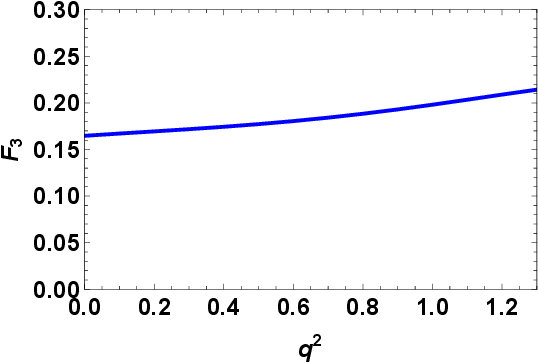}
\includegraphics[totalheight=5cm,width=5.8cm]{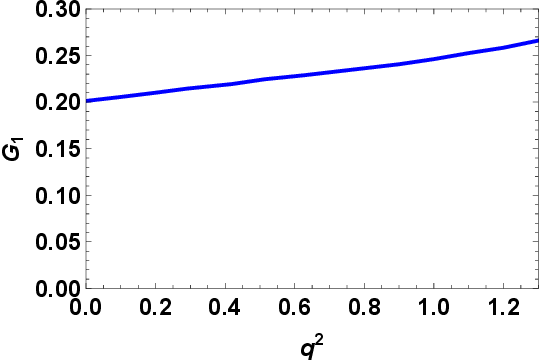}
\includegraphics[totalheight=5cm,width=5.8cm]{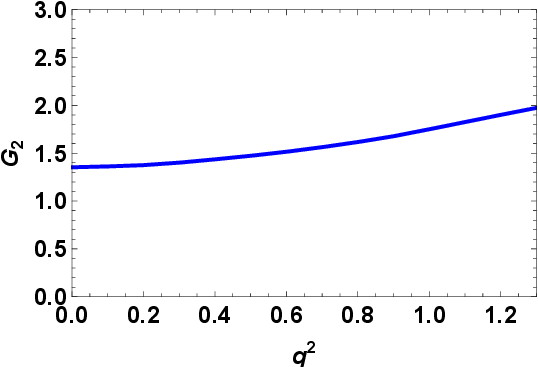}
\includegraphics[totalheight=5cm,width=5.8cm]{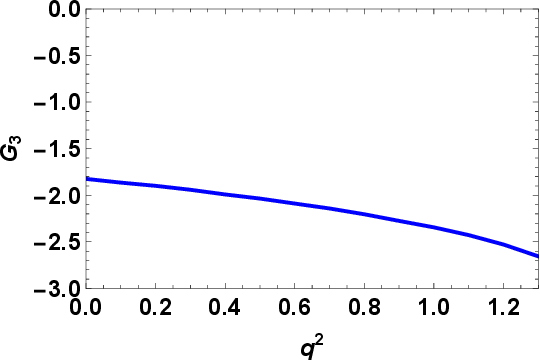}
\caption{The form factors, corresponding to the selected structures, with respect to $q^2$ at the center of values of auxiliary parameters and Ioffe point.}\label{Fig:formfactor1}
\end{figure}

\begin{figure}[h!] 
\includegraphics[totalheight=5cm,width=5.8cm]{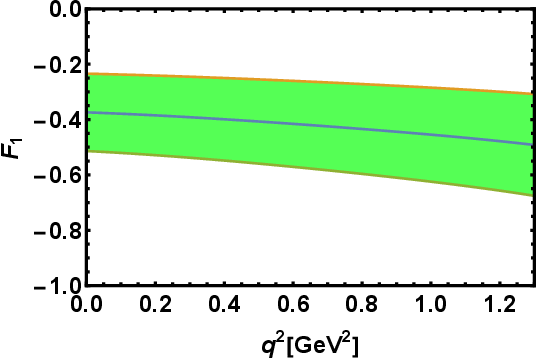}
\includegraphics[totalheight=5cm,width=5.8cm]{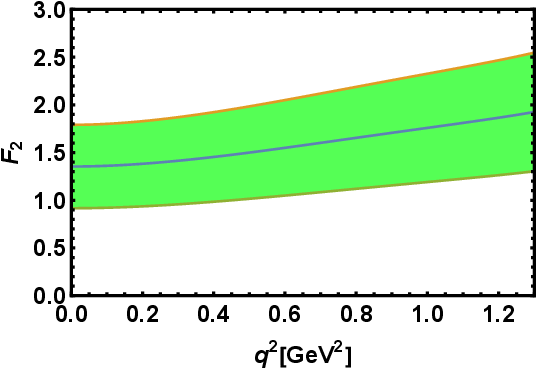}
\includegraphics[totalheight=5cm,width=5.8cm]{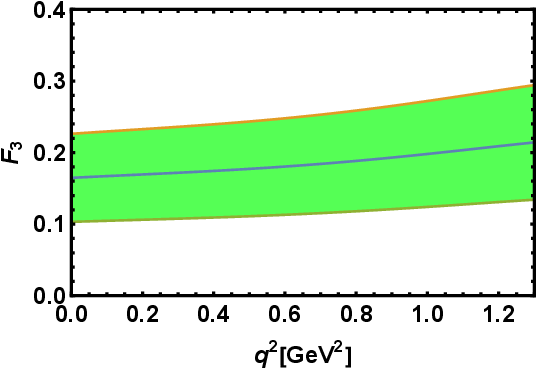}
\includegraphics[totalheight=5cm,width=5.8cm]{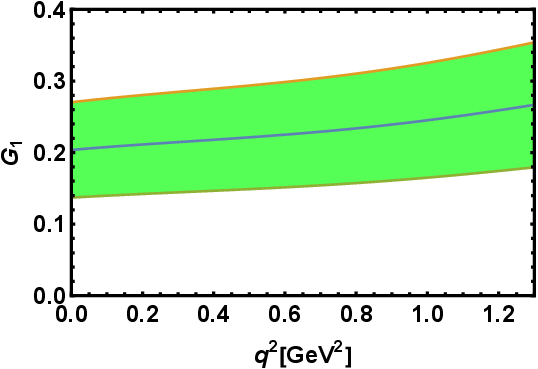}
\includegraphics[totalheight=5cm,width=5.8cm]{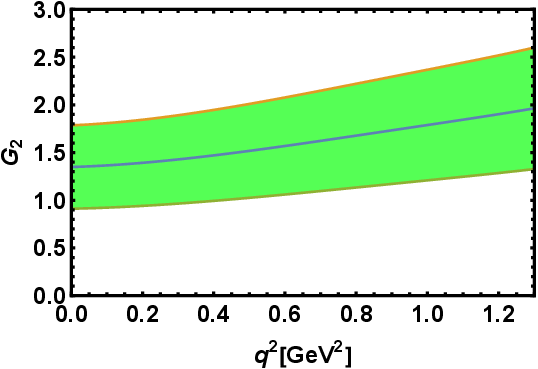}
\includegraphics[totalheight=5cm,width=5.8cm]{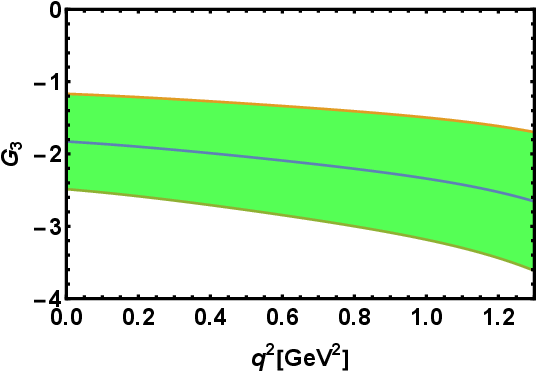}
\caption{The form factors with their errors, corresponding to the selected structures, with respect to  $q^2$ at the center of values of auxiliary parameters and Ioffe point.}\label{Fig:formfactorserror1}
\end{figure}

The QCD sum rules propose several structures for selecting the form factors. The optimal choice are carefully selected by considering the Borel, continuum, and $x$ parameter working intervals, aiming to minimize uncertainties in the results. As previously said, we choose the  $\slashed {p}'\gamma_{\mu}  $, $p_{\mu}\slashed {p'}\slashed {p}$, $p'_{\mu}\slashed {p'}\slashed {p}$, $\gamma_{\mu} \gamma_5$, $p_{\mu}\slashed {p}'\slashed {p}\gamma_5$ and $p'_{\mu}\gamma_5$ structures corresponding to the form factors, $F_1$, $F_2$, $F_3$, $G_1$, $G_2$ and $G_3$, respectively to fulfill the requirements. The uncertainties in the form factors at $q^2=0$ are attributed to the numerical uncertainties in the auxiliary parameter intervals and errors in other input values. Figures \ref{Fig:formfactor1} and \ref{Fig:formfactorserror1} depict the form factors $F_1$, $F_2$, $F_3$, $G_1$, $G_2$ and $G_3$ as functions of $q^2$ at the center of values of $s_0$, $s'_0$, $M^2$, $M'^2$ and the Ioffe point $x = -0.71$ (or $ \beta=-1 $), with and without uncertainties. As expected for the weak transitions, all form factors increase with increasing $q^2$.  It is well-known that, in the baryonic channels, the correlation functions receive contributions from the negative-parity states as well. This point is widely discussed, for instance, in Ref. \cite{Khodjamirian:2011jp} and applied for different baryonic channels in various studies. We have considered the negative parity pollution in our previous paper on the masses and residues of the doubly heavy baryons in Ref. \cite{ShekariTousi:2024mso}. Considering the negative parity (1P) states imposes some systematic uncertainties. In the case of the decay channel under study, the negative parity pollution can be considered for both the initial and final baryonic channels that we ignore in the present study by adjusting the corresponding continuum thresholds. 
 In the next section, we utilize the fitted functions for all six form factors in the range of $m_l^2\leq q^2 \leq (m_{\Xi^{++}_{cc}}-m_{\Xi^{+}_{c}})^2$ to calculate the decay widths and branching ratios.

%%%%%%%%%%%%%%%%%%%%%%%%%%%%%%%%%%%%%%%%%%%%%%%%%%%%%%%%%%%%%%%%%%%%%%%%%%%%%%%%%%%%%%%%%%%%%%%%%%%%%%%%%%%%
\section{Decay Width and Branching Ratio}~\label{Sec4}
%%%%%%%%%%%%%%%%%%%%%%%%%%%%%%%%%%%%%%%%%%%%%%%%%%%%%%%%%%%%%%%%%%%%%%%%%%%%%%%%%%%%%%%%%%%%%%%%%%%%%%%%%%%%
Now, we are ready to present the calculations for the  main physical observables, such as the decay widths and branching ratios associated with  the semileptonic transition  $ \Xi^{++}_{cc}\rightarrow \Xi^+_{c} \bar{\ell}\nu_{\ell}$  in two possible lepton channels  $e^+$ and $\mu ^+$. For this purpose, we utilize the fit functions of the form factors  obtained  in previous section. Using the effective Hamiltonian from Eq. (\ref{Heff}), we proceed to evaluate the decay widths and branching ratios. This calculation involves the use of helicity amplitudes, which are provided for the vector and axial vector currents as follows:

\begin{eqnarray}
	H_{\frac{1}{2},0}^{V} & = & -i\frac{\sqrt{Q_{-}}}{\sqrt{q^{2}}}\left((M_{1}+M_{2})F_{1}-\frac{q^{2}}{M_{1}}F_{2}\right),\;\;\;
	H_{\frac{1}{2},0}^{A} =  -i\frac{\sqrt{Q_{+}}}{\sqrt{q^{2}}}\left((M_{1}-M_{2})G_{1}+\frac{q^{2}}{M_{1}}G_{2}\right),\nonumber \\
	H_{\frac{1}{2},1}^{V} & = & i\sqrt{2Q_{-}}\left(-F_{1}+\frac{M_{1}+M_{2}}{M_{1}}F_{2}\right),\;\;\;
	H_{\frac{1}{2},1}^{A}  =  i\sqrt{2Q_{+}}\left(-G_{1}-\frac{M_{1}-M_{2}}{M_{1}}G_{2}\right),\nonumber \\
	H_{\frac{1}{2},t}^{V} & = & -i\frac{\sqrt{Q_{+}}}{\sqrt{q^{2}}}\left((M_{1}-M_{2})F_{1}+\frac{q^{2}}{M_{1}}F_{3}\right),\;\;\;
	H_{\frac{1}{2},t}^{A} =  -i\frac{\sqrt{Q_{-}}}{\sqrt{q^{2}}}\left((M_{1}+M_{2})G_{1}-\frac{q^{2}}{M_{1}}G_{3}\right),
\end{eqnarray}
where $M_{1}$ and $M_{2}$ are the masses of the initial ($ \Xi^{++}_{cc}$ ) and final ($ \Xi^{+}_{c}$) baryons and  $Q_{\pm}=(M_1\pm M_2)^{2}-q^{2}$. These helicity amplitudes are crucial for determining the decay widths and branching ratios for the two possible lepton channels. The amplitudes for negative helicity are: 
\begin{equation}
	H_{-\lambda_{2},-\lambda_{W}}^{V}=H_{\lambda_{2},\lambda_{W}}^{V}\quad\text{and}\quad H_{-\lambda_{2},-\lambda_{W}}^{A}=-H_{\lambda_{2},\lambda_{W}}^{A},
\end{equation}
where $\lambda_{2}$ and $\lambda_{W}$ define the polarizations of
the final baryon and the $W$ boson, respectively. Finally
the helicity amplitudes for the $V-A$ (vector minus axial vector) current are explained as follows:
\begin{equation}
	H_{\lambda_{2},\lambda_{W}}=H_{\lambda_{2},\lambda_{W}}^{V}-H_{\lambda_{2},\lambda_{W}}^{A}.
\end{equation}

The decay width for  $ \Xi^{++}_{cc}\rightarrow \Xi^+_{c} \bar{\ell}\nu_{\ell}$ channel with the longitudinally and transversely polarized $\bar{\ell}\nu_{\ell}$ pair is calculated as follows:
\begin{align}
	\frac{d\Gamma_{L}}{dq^{2}} & =\frac{G_{F}^{2}|V_{{\rm CKM}}|^{2}q^{2}\ \bold p'\ (1-\hat{m}_{l}^{2})^{2}}{384\pi^{3}M_{1}^{2}}\left((2+\hat{m}_{l}^{2})(|H_{-\frac{1}{2},0}|^{2}+|H_{\frac{1}{2},0}|^{2})+3\hat{m}_{l}^{2}(|H_{-\frac{1}{2},t}|^{2}+|H_{\frac{1}{2},t}|^{2})\right),\label{eq:longi-1}\\
	\frac{d\Gamma_{T}}{dq^{2}} & =\frac{G_{F}^{2}|V_{{\rm CKM}}|^{2}q^{2}\ \bold p'\ (1-\hat{m}_{l}^{2})^{2}(2+\hat{m}_{l}^{2})}{384\pi^{3}M_{1}^{2}}(|H_{\frac{1}{2},1}|^{2}+|H_{-\frac{1}{2},-1}|^{2}),\label{eq:trans-1}
\end{align}

where $\hat{m}_{l}\equiv m_{l}/\sqrt{q^{2}}$, and $\bold p'=\sqrt{Q_{+}Q_{-}}/(2M_{1})$
is the magnitude of three momentum of $ \Xi^{+}_{c}$ in the rest frame of $ \Xi^{++}_{cc}$. By integrating over the $q^{2}$, we find the total decay width as follows: 
\begin{equation}
	\Gamma=\int_{m_l^2}^{(M_{1}-M_{2})^{2}}dq^{2}\frac{d\Gamma}{dq^{2}},
\end{equation}
where
\begin{equation}
	\frac{d\Gamma}{dq^{2}}=\frac{d\Gamma_{L}}{dq^{2}}+\frac{d\Gamma_{T}}{dq^{2}}.
\end{equation}

%%%%%%%%%%%%%%%%%%%
After evaluating the decay width for the two possible lepton channels of $ \Xi^{++}_{cc}\rightarrow \Xi^+_{c} \bar{\ell}\nu_{\ell}$ transition, the average values and their uncertainties are shown  in Table \ref{DECAY}. We also compare our findings with the predictions of  other studies  in the literature. As we can see, our results, considering the uncertainties, are consistent with the predictions made  in Ref. \cite{Shi:2019hbf} for the ($e^+$ and $\mu ^+$) channels. Our results, however,  differ with those of the Ref. \cite{Wang:2017mqp} (presented without uncertainties), considerably. As we previously said, the results of Refs.  \cite{Shi:2019hbf,Wang:2017mqp} are the same for both the leptons.   We further evaluate the branching ratios for the ($e^+$ and $\mu ^+$) channels and show our outcomes  in Table \ref{br}, comparing them with existing theoretical predictions. Our results for branching fractions again are consistent with those of Ref. \cite{Shi:2019hbf} within the presented uncertainties  but differ considerably with the predictions of  Ref. \cite{Wang:2017mqp}. As  is clear from these results, 
our predictions have relatively fewer uncertainties compared with the previous results indicating good  optimization of the results with respect to the variations of the auxiliary parameters.  The obtained results can be checked via  future experiments. 
\begin{table}[bth]
		\centering
\caption{Decay widths (in $\mathrm{GeV}  $) of $ \Xi^{++}_{cc}\rightarrow \Xi^+_{c} \bar{\ell}\nu_{\ell}$ transition for $e^+ $ and $\mu ^+$ channels.}\label{DECAY}
\begin{ruledtabular}
\begin{tabular}{|c|c|c|}	
                 &$\Gamma~[ \Xi^{++}_{cc}\to\Xi^{+}_{c} e^+ \nu_{e}]\times10^{14}$ &  $\Gamma~[ \Xi^{++}_{cc}\to \Xi^{+}_{c} \mu ^+  {\nu}_{\mu }]\times10^{14}$    \\
\hline
This study &$7.20^{+0.77}_{-0.59}$& $7.01^{+0.66}_{-0.35}$     \\
\hline
 Ref. \cite{Wang:2017mqp}  & $12.8$ &$12.8$  \\
\hline
Ref. \cite{Shi:2019hbf}  & $7.72\pm3.70$ &$7.72\pm3.70$  \\
\hline

\end{tabular}
\end{ruledtabular}
\end{table}

\begin{table}[bt]
	\caption{Branching ratios of $ \Xi^{++}_{cc}\rightarrow \Xi^+_{c} \bar{\ell}\nu_{\ell}$ transition for $e^+ $ and $\mu ^+$  channels.}\label{br}
	\begin{ruledtabular}
		\begin{tabular}{|c|c|c|}
		  &$Br~[ \Xi^{++}_{cc}\to\Xi^{+}_{c} e^+ \nu_{e}]\times10^{2}$ &  $Br~[ \Xi^{++}_{cc}\to \Xi^{+}_{c} \mu ^+  {\nu}_{\mu }]\times10^{2}$    \\
			\hline
			This study &$2.80^{+0.49}_{-0.13}$& $2.72^{+0.47}_{-0.13}$     \\
			\hline
			Ref. \cite{Wang:2017mqp}  & $5.84 $ & $5.84 $  \\
			\hline
			Ref. \cite{Shi:2019hbf}  & $3.00\pm1.44$ & $3.00\pm1.44$  \\
			\hline
		\end{tabular}
	\end{ruledtabular}
\end{table}

\section{Conclusion}~\label{Sec5}

After the discovery of the doubly charmed baryon $ \Xi_{cc}$, significant theoretical research initiated focusing on hadron spectroscopy and the decays of the doubly heavy baryons. As the search for these baryons are in agenda of different experiments like LHCb, further investigations of their different features are needed. In this study, we employed QCD sum rules to investigate  the semileptonic weak decay of a doubly heavy baryon into a single heavy one,  $ \Xi^{++}_{cc}\rightarrow \Xi^+_{c} \bar{\ell}\nu_{\ell}$,  in two possible lepton channels. Our analysis encompasses perturbative and nonperturbative condensation contributions up to mass dimension 5.

We computed the form factors of the $ \Xi^{++}_{cc}\rightarrow \Xi^+_{c} \bar{\ell}\nu_{\ell}$ transition entering  the amplitudes described by the the vector and axial vector currents for both  the $e^+$ and $\mu ^+$ channels using the most possible general interpolating currents of the initial and final baryons in full QCD.  After fixing the auxiliary parameters in accordance with the prescriptions of the method,  we derived the fit functions of the relevant six form factors with respect to $q^2$ in the allowed physical region.  We used the obtained results to  estimate the decay widths and branching ratios in both lepton channels.  We also compared the obtained results with the predictions of other theoretical studies. 
 
 The doubly heavy baryons are natural predictions of the quark model. However, as previously said, only the $ \Xi^{++}_{cc}$ has been experimentally identified so far. Considering the progress made in different experiments, like LHCb at CERN, we hope that more ground and excited state doubly heavy baryons will be discovered in the near future. Our results can help experimental groups in this course.  The order of branching fractions obtained for the weak decays considered in this study indicates that these semileptonic decays are easily accessible in the experiments like LHCb. Any experimental data on the physical observables considered and their comparison with the theoretical predictions made can help us not only  to  gain valuable information on the internal structure of the doubly heavy baryons, but also to accurately determine their main parameters like  mass and lifetime.

\section*{APPENDIX: THE EXPRESSIONS OF THE SPECTRAL DENSITIES}
In this appendix, we exhibit the explicit expressions of the spectral densities achieved by our calculations for the structure $\gamma_{\mu} \gamma_5$:
\begin{eqnarray} \label{RhoPert}
&&\rho^{Pert.}_{\gamma_{\mu} \gamma_5}(s,s',q^2)=\int_{0}^{1}du \int_{0}^{1-u}dv \int_{0}^{1-u-v}dz~\frac{\sqrt{3}  }{256 A H^3 J^6 \pi^4} \notag\\
&& \bigg\{-D_1^2 J^4 \bigg[B (11 + 7 \beta) H (H - 2 u) v + 
z \bigg(-\big((11 + 7 \beta) C H (1 + 2 u)\big) - 
2 u z \big(18 u + 7 (-1 + \beta) z\big) + 
H \big(v (-40 - 18 u \notag\\
&&+ 7 \beta (-2 + v) + 29 v) + \big(-22 + 3 u + 
29 v + 7 \beta (-2 - 3 u + v)\big) z + (11 + 
7 \beta) z^2\big)\bigg)\bigg] - 
2 A H^2 v^2 z \bigg[H J^2 m_c^2 \big(A \beta s' \notag\\
&&+ 4 (-q^2 + s) u - 
s' (5 H + u + 5 z)\big) + 
s' z \bigg\{-\big[(-H + u - z) \big((5 + \beta) H s' + 2 (2 + \beta) s' u + 
q^2 (u - \beta u) + (5 + \beta) s' z\big)\notag\\
&& \big(B H v + 
z (C H + 2 H v + H z + 2 u z)\big)\big] - 
s \bigg[B H \big(3 (1 + \beta) H^2 + 2 (2 + \beta) H u - 
6 (2 + \beta) u^2\big) v + 
z \bigg(H \big(-3 (1 + \beta) \notag\\
&&+ (7 + 5 \beta) u + 4 (2 + \beta) u^2 - 
6 (2 + \beta) u^3 + \big[15 - 7 u (4 + u) + 
\beta \big(15 + (-20 + u) u\big)\big] v + 
3 (-7 - 7 \beta + 7 u + 5 \beta u) v^2 \notag\\
&&+ 
9 (1 + \beta) v^3\big) + \big(6 (-1 + \beta) u^3 + (7 + 
5 \beta) H u (-2 + 3 v) + 
3 (1 + \beta) H^2 (-3 + 4 v)\big) z + \big(9 (1 + 
\beta) H^2 + (7 + 5 \beta) H u \notag\\
&&+ 
4 (2 + \beta) u^2\big) z^2 + 3 (1 + \beta) H z^3\bigg)\bigg]\bigg\}\bigg] - 
2 D_1 H J^2 v \bigg\{2 A (-4 + \beta) H J^2 m_c^2 + 
u z \bigg[q^2 \bigg\{B H v \big(-9 + 13 u + \beta (3 + 13 u - 3 v) \notag\\
&&+ 9 v\big) + 
z \bigg(C H (3 (-3 + \beta) + 13 (1 + \beta) u) + 
H v \big(4 (-9 + 3 \beta + 11 u + 5 \beta u) - 
9 (-3 + \beta) v\big) + (-6 (-3 + \beta) - 
8 (5 + \beta) u \notag\\
&&+ 2 (13 + 15 \beta) u^2 + 
15 (-3 + \beta) v + 8 (5 + \beta) u v - 
9 (-3 + \beta) v^2) z + \big(9 (H + 2 u) + 
\beta (3 - 2 u - 3 v)\big) z^2\bigg)\bigg\} + 
s \bigg(B H v (5 + \beta \notag\\
&&+ 2 u + 2 \beta u - (5 + \beta) v) + 
z \big[C H \big(5 + \beta + 2 (1 + \beta) u\big) + 
H v \big(24 - 4 (4 + 3 \beta) u + (-19 + \beta) v\big) + \bigg(-10 - 
2 \beta + 21 u \notag\\
&&- 3 \beta u - 16 u^2 - 
28 \beta u^2 + \big(29 + \beta + 
3 (-7 + \beta) u\big) v + (-19 + \beta) v^2\bigg) z + \big(5 + 
\beta - 18 u + 2 \beta u - (5 + \beta) v\big) z^2\big]\bigg)\bigg] + 
2 s' z \notag\\
&&\bigg[2 B H v \big(5 + \beta + 5 u + \beta u - 11 u^2 - 
6 \beta u^2 - (5 + \beta) (2 + u) v + (5 + \beta) v^2\big) + 
z \bigg(2 H^3 \big(-5 - \beta + 3 (7 + \beta) v \big) \notag\\
&&+ 
H^2 \big(-3 (-1 + \beta) u v - 6 (5 + \beta) z + 
8 (8 + \beta) v z + 6 \beta z^2\big) + 
H \bigg(-u^2 \big(-32 + 22 u + 63 v + \beta (-14 + 12 u + 29 v)\big) \notag\\
&&- 
u \big(21 + 52 u + \beta (3 + 20 u) - 24 v\big) z + 
6 (-5 + 7 u + 7 v) z^2\bigg) + 
z \big[-5 (9 + 5 \beta) u^3 - 2 (10 + 3 \beta) u^2 z + 
2 (5 + \beta) (-1 + v) z^2 \notag\\
&&+ 
3 u z \big(2 \beta (-1 + v) + (7 + \beta) z\big)\big]\bigg)\bigg]\bigg\} + 
\beta \bigg[-D_1^2 J^4 \bigg(B (7 + 11 \beta) H (H - 2 u) v + 
H \bigg(-\big((7 + 11 \beta) C (1 + 2 u)\big) + 7 (-2 + v) v \notag\\
&&+ 
\beta v (-40 - 18 u + 29 v)\bigg) z + \big(-36 \beta u^2 + 
7 H (-2 - 3 u + v) + 
\beta H (-22 + 3 u + 29 v)\big) z^2 + \big((7 + 11 \beta) H + 
14 (-1 + \beta) u\big) \notag\\
&&z^3\bigg) + 
2 D_1 H J^2 v \bigg[2 A (-4 + \beta) H J^2 m_c^2 - 
z \bigg(4 B H s' v (1 + 5 \beta + u + 5 \beta u - 6 u^2 - 
11 \beta u^2 - (1 + 5 \beta) (2 + u) v + (1 + 
5 \beta) v^2) \notag\\
&&+ 
q^2 u \bigg[B H v \big(3 + 13 u - 3 v + \beta (-9 + 13 u + 9 v)\big) + 
z \bigg(C H \big(3 + 13 u + \beta (-9 + 13 u)\big) + 
H v \big(12 + 20 u - 9 v + 
\beta (-36 + 44 u \notag\\
&&+ 27 v)\big) + \big(-6 + 8 H u + 
30 u^2 + 26 \beta u^2 + 15 v - 9 v^2 + 
\beta H (-18 + 40 u + 27 v)\big) z + \big(3 - 2 u + 
9 \beta (H + 2 u) - 3 v\big) z^2\bigg)\bigg] \notag\\
&&+ 
s u \bigg[B H \big(-H + 2 u + \beta (5 + 2 u - 5 v)\big) v + 
z \big\{C H \big(1 + 5 \beta + 2 (1 + \beta) u\big) + 
H v \big(-12 u + \beta (24 - 16 u - 19 v) + v\big) + \bigg(-2\notag\\
&& + 
3 H u - 28 u^2 + v + v^2 - 
\beta \big(16 u^2 + H (-10 + 21 u + 19 v)\big)\bigg) z - \big(H - 
2 u + \beta (-5 + 18 u + 5 v)\big) z^2\big\}\bigg] + 
2 s' z \big[2 H^3 \big(-1 + 3 v \notag\\
&&+ \beta (-5 + 21 v)\big) + 
z \big(-5 (5 + 9 \beta) u^3 - 
2 (3 + 10 \beta) u^2 z + (-2 + 3 u + 21 \beta u + 
2 v) z^2\big) + 
H \big[-u^2 \big(-14 + 12 u + 29 v + \beta (-32 \notag\\
&&\notag\\
&&+ 22 u + 63 v)\big) -
u \big(3 + 20 u + \beta (21 + 52 u - 24 v)\big) z + 
6 \big(u + \beta (-5 + 7 u + 7 v)\big) z^2 + 10 \beta z^3\big] + 
H^2 \big(3 (-1 + \beta) u v \notag\\
&&\notag\\
&&
+ 
2 z (-3 - 15 \beta + 4 v + 32 \beta v + 3 z)\big)\big]\bigg)\bigg]+ 
2 A H^2 v^2 z \bigg(H J^2 m_c^2 \big(A \beta s' + 4 (-q^2 + s) u - 
s' (5 H + u + 5 z)\big) + 
s' z \big\{(-H + u \notag\\
&&- z) \big[(-1 + \beta) q^2 u + 
s' \big(H + 2 u + 4 \beta u + 5 \beta (F + v) + z\big)\big] \big(B H v + 
z (C H + 2 H v + H z + 2 u z)\big) + 
s \big[B H (3 (1 + \beta) H^2 \notag\\
&&- 6 (1 + 2 \beta) u^2 + 
2 H \big(u + 2 \beta u)\big) v + 
z \big(C H \big[3 - 2 u (1 + 3 u) + 
\beta \big(3 - 4 u (1 + 3 u) + 21 v^2\big)\big] + 
3 (1 + \beta) H^2 z (-3 + 4 v + 3 z) \notag\\
&&
+2 u^2 z (3 u - 3 \beta u + 2 z + 4 \beta z) + 
H \{3 (-7 + 5 u) v^2 + 
9 (1 + \beta) v^3 + (5 + 7 \beta) u (-2 + z) z + 
3 (1 + \beta) z^3 + 
v [15 + \beta [15 \notag\\
&&- 7 u (4 + u - 3 z)] + 
u (-20 + u + 15 z)]\}\big)\big]\big\}\bigg)\bigg] \bigg\} \Theta[D_1(s,s^{\prime},q^2)], 
\end{eqnarray}

\begin{eqnarray} \label{Rho3}
&&\rho^3_ {\gamma_{\mu} \gamma_5}(s,s',q^2)=
\int_{0}^{1}du \int_{0}^{1-u}dv ~\frac{ 1 }{32 \sqrt{3}  C^3 \pi^2} \notag\\
&&\bigg[-\bigg(\big(11 + \beta (14 + 11 \beta)\big) m_c \langle \bar{s} s\rangle \Theta[L(s,s^{\prime},q^2)]\bigg) + 
2 (1 + \beta - 2 \beta^2) C^3 m_c \langle \bar{u} u\rangle \Theta[L'(s,s^{\prime},q^2)]\bigg],
\end{eqnarray}

\begin{eqnarray}\label{Rho4}
&&\rho^4_{\gamma_{\mu} \gamma_5}(s,s',q^2)=\int_{0}^{1}du \int_{0}^{1-u}dv \int_{0}^{1-u-v}dz\frac{1}{1024 H^5 J^4 \pi^2} \langle0|\frac{1}{\pi}\alpha_s G^2|0\rangle \bigg\{2 H^5 \bigg[2 (1 + \beta^2) F^2 u z - 
2 \big(19 + \beta (12 + 11 \beta)\big) \notag\\
&&u^3 (v + z) + (1 + \beta)^2 v^2 (H +
z) (H + 9 z) + 
u v^2 (2 (1 + \beta^2) (-2 + v) + (35 + 
3 \beta (14 + 9 \beta)) z) + 
F \bigg(-4 \big(9 + \beta (6 + 5 \beta)\big)  \notag\\
&&u^2 z+ (1 + \beta)^2 v z (H + 
z) + 2 u v \big(-1 - 
\beta^2 + [13 + 3 \beta (4 + 3 \beta)] z\big)\bigg) - 
2 u^2 v \bigg(-18 + 19 v + 27 z + 
\beta \big(2 (-6 + 7 v + 7 z)\notag\\
&& + 
\beta (-10 + 11 v + 15 z)\big)\bigg)\bigg]  + 
z \bigg[H^4 v^2 \bigg(4 u (-1 + 2 u - 5 v) + 2 H v + 
2 \beta^2 [2 u (-1 + 2 u - 5 v) + H v] \notag\\
&&- 
\beta \big(14 + 3 (-13 + u) u - 26 v + 55 u v + 12 v^2 + 
2 H (-7 + 6 v)\big)\bigg) + 
H^4 v \bigg(-6 \big(1 + \beta (6 + \beta)\big) + 2 (1 + \beta (45 + \beta)) u \notag\\
&&+ 
2 \big(8 + \beta (-3 + 8 \beta)\big) u^2 - \bigg(2 + 50 u + 
\beta \big(-92 + 201 u + \beta (2 + 50 u)\big)\bigg) v+ 
16 \big(1 + (-3 + \beta) \beta \big) v^2\bigg) z  \notag\\
&&+ 
H^2 \bigg[8 u^2 \big(1 + 2 (-2 + v) v\big) - 4 u (1 + v - 17 v^2 + 9 v^3) + 
4 H v \big(-1 + v (-3 + 2 v)\big) + 
4 \beta^2 \bigg(u^2 \big(2 + 4 (-2 + v) v\big)  \notag\\
&&- 
u (1 + v - 17 v^2 + 9 v^3) + H v \big(-1 + v (-3 + 2 v)\big)\bigg) + 
\beta \bigg(u \big(39 - 3 u \big(1 + 2 (-2 + v) v\big) + 
v (-369 + 574 v - 228 v^2)\big)  \notag\\
&&- 
4 H \big[7 + v \big(49 + v (-71 + 31 v)\big)\big]\bigg)\bigg] z^2 + 
H^2 \bigg(\beta (96 + 6 u^2 + 8 v \big(-44 + (49 - 18 v) v\big) + 
u (-157 + 346 v - 78 v^2)) \notag\\
&& + 
4 \bigg(-4 u^2 + v \big(3 + (-6 + v) v\big) + 2 u \big(1 + v (4 + v)\big)\bigg) + 
4 \beta^2 \big[-4 u^2 + v \big(3 + (-6 + v) v\big) + 
2 u \big(1 + v (4 + v)\big)\big]\bigg) z^3  \notag\\
&&- 
2 H \big[-2 u + 
\beta \big(-27 + 27 H + (59 - 2 \beta) u\big) + \bigg(5 H + 
\beta \big(66 + (-39 + 5 \beta) H\big) + 4 u + 
4 \beta (10 + \beta) u  \notag\\
&&+ (8 + 
\beta \big(-3 + 8 \beta)\big) u^2\bigg) v + \bigg(11 H - 26 u + 
\beta \big(-56 + (17 + 11 \beta) H - (115 + 26 \beta) u\big)\bigg) v^2 + 
17 \beta v^3\big] z^4  \notag\\
&&+ 
2 H \bigg(-8 (u - 2 v) (u - v) + 8 v + 
8 \beta^2 \big(-\big[(u - 2 v) (u - v)\big] + v\big) + 
\beta \big(-8 + 3 u^2 + 12 H v + 
u (-79 + 174 v)\big)\bigg) z^5  \notag\\
&&+ \bigg(-8 (u^2 + 3 H v - 3 u v) - 
8 \beta^2 (u^2 + 3 H v - 3 u v) + 
\beta \big(60 H^2 + 3 (-79 + u) u + 253 u v\big)\bigg) z^6 + \bigg(\beta (-48 + 
79 u)  \notag\\
&&- 8 \big(1 + (-5 + \beta) \beta\big) v\bigg) z^7 + 
12 \beta z^8\bigg]\bigg\} \Theta[D_2(s,s^{\prime},q^2)],
\end{eqnarray}

\begin{equation}\label{Rho5}
	\rho^5_{\gamma_{\mu} \gamma_5}(s,s',q^2)=0,
\end{equation}
%%%%%
where,
\begin{eqnarray}\label{L}
&&D_1(s,s^{\prime},q^2)=\frac{-H}{J^2} \bigg(-q^2 u (v + z) A + 
	m_c^2 (u + v) (v^2 + v F + F z) + 
	v z (s u - s' A)\bigg),
\\
&&D_2(s,s^{\prime},q^2)=\frac{-H }{J^2} \bigg(-q^2 u (v + z) A + 
	m_c^2 (u + v) J + 
	v z (s u - s' A)\bigg),
\\	
&&L=\frac{1}{H^2} \bigg(-s C u + (-q^2 + s + s') C u + m_c^2 C (u + v) - 
s' (u^2 + u H + H v)\bigg),
\\
&&L'=-s C u + (-q^2 + s + s') u B - m_c^2 (u + v) + 
s' (u - u^2 + v - 2 u v - v^2),
\end{eqnarray}
and $\Theta[...]$ stands for the unit step function.
We have defined
\begin{eqnarray}\label{dL}
&&J=v^2 + v F + F z,\notag
\\
&&A=-1 + u + v + z,\notag
\\
&&B=-1 + u + v,\notag
\\
&&C=-1 + u,\notag
\\
&&H=-1 + v,
\\
&&F=-1 + z.\notag
\end{eqnarray}
%

%%%%%%%%%%%%%%%%%%%%%%%%%%%%%%%%%%%%%%%%%%%%%%%%%%%%%%%%%%%
%%%%%%%%%%%%%%%%%%%%%%%%%%%%%%%%%%%%%%%%%%%%%%%%%%%%%%%%%%%%%%%%%%%%%%%%%%%%%%%%%%%%%%%%%%%%%%%%%%%%%%%%%%%%


\begin{thebibliography}{99}




%\cite{GellMann:1964nj}
\bibitem{GellMann:1964nj}
M.~Gell-Mann,
``A Schematic Model of Baryons and Mesons,''
\href{https://www.sciencedirect.com/science/article/abs/pii/S0031916364920013?via%3Dihub}{Phys. Lett. \textbf{8}, 214-215 (1964)}.
%doi:10.1016/S0031-9163(64)92001-3
%4064 citations counted in INSPIRE as of 02 Dec 2023

%\cite{SELEX:2002wqn}
\bibitem{SELEX:2002wqn}
M.~Mattson \textit{et al.} [SELEX],
``First Observation of the Doubly Charmed Baryon $\Xi^+_{cc}$,''
\href{https://doi.org/10.1103/PhysRevLett.89.112001}{Phys. Rev. Lett. \textbf{89}, 112001 (2002)},
\href{https://arxiv.org/abs/hep-ex/0208014}{[arXiv:hep-ex/0208014 [hep-ex]]}.
%497 citations counted in INSPIRE as of 22 Nov 2023

%\cite{SELEX:2004lln}
\bibitem{SELEX:2004lln}
A.~Ocherashvili \textit{et al.} [SELEX],
``Confirmation of the double charm baryon Xi+(cc)(3520) via its decay to p D+ K-,''
\href{https://doi.org/10.1016/j.physletb.2005.09.043}{Phys. Lett. B \textbf{628}, 18-24 (2005)},
\href{https://arxiv.org/abs/hep-ex/0406033}{[arXiv:hep-ex/0406033 [hep-ex]]}.
%378 citations counted in INSPIRE as of 22 Nov 2023

%\cite{LHCb:2017iph}
\bibitem{LHCb:2017iph}
R.~Aaij \textit{et al.} [LHCb],
``Observation of the doubly charmed baryon $\Xi_{cc}^{++}$,''
\href{https://doi.org/10.1103/PhysRevLett.119.112001}{Phys. Rev. Lett. \textbf{119}, no.11, 112001 (2017)},
\href{https://arxiv.org/abs/1707.01621}{[arXiv:1707.01621 [hep-ex]]}.
%533 citations counted in INSPIRE as of 01 Dec 2023


%\cite{LHCb:2018pcs}
\bibitem{LHCb:2018pcs}
R.~Aaij \textit{et al.} [LHCb],
``First Observation of the Doubly Charmed Baryon Decay $\Xi_{cc}^{++}\rightarrow \Xi_{c}^{+}\pi^{+}$,''
\href{https://doi.org/10.1103/PhysRevLett.121.162002}{Phys. Rev. Lett. \textbf{121}, no.16, 162002 (2018)},
\href{https://arxiv.org/abs/1807.01919}{[arXiv:1807.01919 [hep-ex]]}.
%163 citations counted in INSPIRE as of 01 Dec 2023

%\cite{LHCb:2019gqy}
\bibitem{LHCb:2019gqy}
R.~Aaij \textit{et al.} [LHCb],
``Search for the doubly charmed baryon $\Xi_{cc}^+$,''
\href{https://doi.org/10.1007/s11433-019-1471-8}{Sci. China Phys. Mech. Astron. \textbf{63}, no.2, 221062 (2020)},
\href{https://arxiv.org/abs/1909.12273}{[arXiv:1909.12273 [hep-ex]]}.
%58 citations counted in INSPIRE as of 03 Feb 2024



%\cite{LHCb:2021eaf}
\bibitem{LHCb:2021eaf}
R.~Aaij \textit{et al.} [LHCb],
``Search for the doubly charmed baryon $ {\varXi}_{cc}^{+} $ in the $ {\varXi}_c^{+}{\pi}^{-}{\pi}^{+} $ final state,''
\href{https://doi.org/10.1007/JHEP12(2021)107}{JHEP \textbf{12}, 107 (2021)},
\href{https://arxiv.org/abs/2109.07292}{[arXiv:2109.07292 [hep-ex]]}.
%26 citations counted in INSPIRE as of 03 Feb 2024



%\cite{LHCb:2021xba}
\bibitem{LHCb:2021xba}
R.~Aaij \textit{et al.} [LHCb],
``Search for the doubly heavy baryons $\Omega^0_{bc}$ and $\Xi^0_{bc}$ decaying to $\Lambda^+_c \pi^-$ and $\Xi^+_c \pi^-$,''
\href{https://doi.org/10.1088/1674-1137/ac0c70}{Chin. Phys. C \textbf{45}, no.9, 093002 (2021)},
\href{https://arxiv.org/abs/2104.04759}{[arXiv:2104.04759 [hep-ex]]}.
%33 citations counted in INSPIRE as of 06 Jan 2024

%\cite{ShekariTousi:2024mso}
\bibitem{ShekariTousi:2024mso}
M.~Shekari Tousi and K.~Azizi,
``Properties of doubly heavy spin-1/2 baryons: The ground and excited states,''
\href{https://journals.aps.org/prd/abstract/10.1103/PhysRevD.109.054005}{Phys. Rev. D \textbf{109}, no.5, 054005 (2024)},
\href{https://arxiv.org/abs/2401.07151}{[arXiv:2401.07151 [hep-ph]]}.
%0 citations counted in INSPIRE as of 14 Jun 2024

\bibitem{Aliev:2012ru}
T.~M.~Aliev, K.~Azizi and M.~Savci,
``Doubly Heavy Spin--1/2 Baryon Spectrum in QCD,''
\href{https://doi.org/10.1016/j.nuclphysa.2012.09.009}{Nucl. Phys. A \textbf{895}, 59-70 (2012)},
\href{https://arxiv.org/abs/1205.2873}{[arXiv:1205.2873 [hep-ph]]}.
%69 citations counted in INSPIRE as of 22 Nov 2023


\cite{Aliev:2012iv}
\bibitem{Aliev:2012iv}
T.~M.~Aliev, K.~Azizi and M.~Savci,
``The masses and residues of doubly heavy spin-3/2 baryons,''
\href{https://doi.org/10.1088/0954-3899/40/6/065003}{J. Phys. G \textbf{40}, 065003 (2013)},
\href{https://arxiv.org/abs/1208.1976}{[arXiv:1208.1976 [hep-ph]]}.
%60 citations counted in INSPIRE as of 08 Nov 2023

%\cite{Aliev:2019lvd}
\bibitem{Aliev:2019lvd}
T.~M.~Aliev and S.~Bilmis,
``The mass and residues of radially and orbitally excited doubly heavy baryons in QCD,''
\href{https://doi.org/10.1016/j.nuclphysa.2019.02.001}{Nucl. Phys. A \textbf{984}, 99-111 (2019)},
\href{https://arxiv.org/abs/1904.11279}{[arXiv:1904.11279 [hep-ph]]}.
%9 citations counted in INSPIRE as of 08 Nov 2023



%\cite{Aliyev:2022rrf}
\bibitem{Aliyev:2022rrf}
T. M.~Aliev and S.~Bilmi\c{s},
``Properties of doubly heavy baryons in QCD,''
\href{https://doi.org/10.3906/fiz-2202-17}{Turk. J. Phys. \textbf{46}, no.1, 1-26 (2022)},
\href{https://arxiv.org/abs/2203.02965}{[arXiv:2203.02965 [hep-ph]]}.
%5 citations counted in INSPIRE as of 08 Nov 2023


%\cite{Lu:2017meb}
\bibitem{Lu:2017meb}
Q.~F.~L\"u, K.~L.~Wang, L.~Y.~Xiao and X.~H.~Zhong,
``Mass spectra and radiative transitions of doubly heavy baryons in a relativized quark model,''
\href{https://doi.org/10.1103/PhysRevD.96.114006}{Phys. Rev. D \textbf{96}, no.11, 114006 (2017)},
\href{https://arxiv.org/abs/1708.04468}{[arXiv:1708.04468 [hep-ph]]}.
%81 citations counted in INSPIRE as of 08 Nov 2023


%\cite{Wang:2018lhz}
\bibitem{Wang:2018lhz}
Z.~G.~Wang,
``Analysis of the doubly heavy baryon states and pentaquark states with QCD sum rules,''
\href{https://doi.org/10.1140/epjc/s10052-018-6300-4}{Eur. Phys. J. C \textbf{78}, no.10, 826 (2018)},
\href{https://arxiv.org/abs/1808.09820}{[arXiv:1808.09820 [hep-ph]]}.
%49 citations counted in INSPIRE as of 02 Feb 2024




%\cite{Ebert:2002ig}
\bibitem{Ebert:2002ig}
D.~Ebert, R.~N.~Faustov, V.~O.~Galkin and A.~P.~Martynenko,
``Mass spectra of doubly heavy baryons in the relativistic quark model,''
\href{https://doi.org/10.1103/PhysRevD.66.014008}{Phys. Rev. D \textbf{66}, 014008 (2002)},
\href{https://arxiv.org/abs/hep-ph/0201217}{[arXiv:hep-ph/0201217 [hep-ph]]}.
%270 citations counted in INSPIRE as of 22 Nov 2023

%\cite{Zhang:2008rt}
\bibitem{Zhang:2008rt}
J.~R.~Zhang and M.~Q.~Huang,
``Doubly heavy baryons in QCD sum rules,''
\href{https://doi.org/10.1103/PhysRevD.78.094007}{Phys. Rev. D \textbf{78}, 094007 (2008)},
\href{https://arxiv.org/abs/0810.5396}{[arXiv:0810.5396 [hep-ph]]}.
%103 citations counted in INSPIRE as of 08 Nov 2023

%\cite{Wang:2010hs}
\bibitem{Wang:2010hs}
Z.~G.~Wang,
``Analysis of the ${1\over 2}^+$ doubly heavy baryon states with QCD sum rules,''
\href{https://doi.org/10.1140/epja/i2010-11004-3}{Eur. Phys. J. A \textbf{45}, 267-274 (2010)},
\href{https://arxiv.org/abs/1001.4693}{[arXiv:1001.4693 [hep-ph]]}.
%101 citations counted in INSPIRE as of 11 Dec 2023

%\cite{Rahmani:2020pol}
\bibitem{Rahmani:2020pol}
S.~Rahmani, H.~Hassanabadi and H.~Sobhani,
``Mass and decay properties of double heavy baryons with a phenomenological potential model,''
\href{https://doi.org/10.1140/epjc/s10052-020-7867-0}{Eur. Phys. J. C \textbf{80}, no.4, 312 (2020)}.
%14 citations counted in INSPIRE as of 27 Nov 2023


%\cite{Yao:2018ifh}
\bibitem{Yao:2018ifh}
D.~L.~Yao,
``Masses and sigma terms of doubly charmed baryons up to $\mathcal{O}(p^4)$ in manifestly Lorentz-invariant baryon chiral perturbation theory,''
\href{https://doi.org/10.1103/PhysRevD.97.034012}{Phys. Rev. D \textbf{97}, no.3, 034012 (2018)},
\href{https://arxiv.org/abs/1801.09462}{[arXiv:1801.09462 [hep-ph]]}.
%33 citations counted in INSPIRE as of 28 Nov 2023




%\cite{Padmanath:2019ybu}
\bibitem{Padmanath:2019ybu}
M.~Padmanath,
``Heavy baryon spectroscopy from lattice QCD,''
\href{https://arxiv.org/abs/1905.10168}{[arXiv:1905.10168 [hep-lat]]}.
%7 citations counted in INSPIRE as of 10 Dec 2023


%\cite{Brown:2014ena}
\bibitem{Brown:2014ena}
Z.~S.~Brown, W.~Detmold, S.~Meinel and K.~Orginos,
``Charmed bottom baryon spectroscopy from lattice QCD,''
\href{https://doi.org/10.1103/PhysRevD.90.094507}{Phys. Rev. D \textbf{90}, no.9, 094507 (2014)},
\href{https://arxiv.org/abs/1409.0497}{[arXiv:1409.0497 [hep-lat]]}.
%240 citations counted in INSPIRE as of 10 Dec 2023


%\cite{Shah:2017liu}
\bibitem{Shah:2017liu}
Z.~Shah and A.~K.~Rai,
``Excited state mass spectra of doubly heavy $\Xi$ baryons,''
\href{https://doi.org/10.1140/epjc/s10052-017-4688-x}{Eur. Phys. J. C \textbf{77}, no.2, 129 (2017)},
\href{https://arxiv.org/abs/1702.02726}{[arXiv:1702.02726 [hep-ph]]}.
%66 citations counted in INSPIRE as of 08 Nov 2023


%\cite{Giannuzzi:2009gh}
\bibitem{Giannuzzi:2009gh}
F.~Giannuzzi,
``Doubly heavy baryons in a Salpeter model with AdS/QCD inspired potential,''
\href{https://doi.org/10.1103/PhysRevD.79.094002}{Phys. Rev. D \textbf{79}, 094002 (2009)},
\href{https://arxiv.org/abs/0902.4624}{[arXiv:0902.4624 [hep-ph]]}.
%56 citations counted in INSPIRE as of 10 Dec 2023





%\cite{Shah:2016vmd}
\bibitem{Shah:2016vmd}
Z.~Shah, K.~Thakkar and A.~K.~Rai,
``Excited State Mass spectra of doubly heavy baryons $\Omega_{cc}$, $\Omega_{bb}$ and $\Omega_{bc}$,''
\href{https://doi.org/10.1140/epjc/s10052-016-4379-z}{Eur. Phys. J. C \textbf{76}, no.10, 530 (2016)},
\href{https://arxiv.org/abs/1609.03030}{[arXiv:1609.03030 [hep-ph]]}.
%78 citations counted in INSPIRE as of 27 Nov 2023




%\cite{Valcarce:2008dr}
\bibitem{Valcarce:2008dr}
A.~Valcarce, H.~Garcilazo and J.~Vijande,
``Towards an understanding of heavy baryon spectroscopy,''
\href{https://doi.org/10.1140/epja/i2008-10616-4}{Eur. Phys. J. A \textbf{37}, 217-225 (2008)},
\href{https://arxiv.org/abs/0807.2973}{[arXiv:0807.2973 [hep-ph]]}.
%161 citations counted in INSPIRE as of 09 Nov 2023

%\cite{Wang:2010it}
\bibitem{Wang:2010it}
Z.~G.~Wang,
``Analysis of the ${1/2^-}$ and ${3/2^-}$ heavy and doubly heavy baryon states with QCD sum rules,''
\href{https://doi.org/10.1140/epja/i2011-11081-8}{Eur. Phys. J. A \textbf{47}, 81 (2011)},
\href{https://arxiv.org/abs/1003.2838}{[arXiv:1003.2838 [hep-ph]]}.
%79 citations counted in INSPIRE as of 08 Nov 2023


%\cite{Yoshida:2015tia}
\bibitem{Yoshida:2015tia}
T.~Yoshida, E.~Hiyama, A.~Hosaka, M.~Oka and K.~Sadato,
``Spectrum of heavy baryons in the quark model,''
\href{https://doi.org/10.1103/PhysRevD.92.114029}{Phys. Rev. D \textbf{92}, no.11, 114029 (2015)},
\href{https://arxiv.org/abs/1510.01067}{[arXiv:1510.01067 [hep-ph]]}.
%202 citations counted in INSPIRE as of 01 Dec 2023

%\cite{Ortiz-Pacheco:2023kjn}
\bibitem{Ortiz-Pacheco:2023kjn}
E.~Ortiz-Pacheco and R.~Bijker,
``Masses and radiative decay widths of S- and P-wave singly, doubly, and triply heavy charm and bottom baryons,''
\href{https://doi.org/10.1103/PhysRevD.108.054014}{Phys. Rev. D \textbf{108}, no.5, 054014 (2023)},
\href{https://arxiv.org/abs/2307.04939}{[arXiv:2307.04939 [hep-ph]]}.
%6 citations counted in INSPIRE as of 09 Jan 2024


%\cite{Aliev:2012nn}
\bibitem{Aliev:2012nn}
T.~M.~Aliev, K.~Azizi and M.~Savc\i{},
``Mixing angle of doubly heavy baryons in QCD,''
\href{https://doi.org/10.1016/j.physletb.2012.07.033}{Phys. Lett. B \textbf{715}, 149-151 (2012)},
\href{https://arxiv.org/abs/1205.6320}{[arXiv:1205.6320 [hep-ph]]}.


%\cite{Qiu:2020omj}
\bibitem{Qiu:2020omj}
P.~C.~Qiu and D.~L.~Yao,
``Chiral effective Lagrangian for doubly charmed baryons up to $\mathcal{O}(q^4)$,''
\href{https://doi.org/10.1103/PhysRevD.103.034006}{Phys. Rev. D \textbf{103}, no.3, 034006 (2021)},
\href{https://arxiv.org/abs/2012.11117}{[arXiv:2012.11117 [hep-ph]]}.
%8 citations counted in INSPIRE as of 08 Nov 2023




%\cite{Alrebdi:2020rev}
\bibitem{Alrebdi:2020rev}
H.~I.~Alrebdi, T.~M.~Aliev and K.~\c{S}im\c{s}ek,
``Determination of the strong vertices of doubly heavy baryons with pseudoscalar mesons in QCD,''
\href{https://doi.org/10.1103/PhysRevD.102.074007}{Phys. Rev. D \textbf{102}, no.7, 074007 (2020)},
\href{https://arxiv.org/abs/2008.05098}{[arXiv:2008.05098 [hep-ph]]}.
%11 citations counted in INSPIRE as of 08 Nov 2023







%\cite{Olamaei:2021hjd}
\bibitem{Olamaei:2021hjd}
A.~R.~Olamaei, K.~Azizi and S.~Rostami,
``Strong vertices of doubly heavy spin-3/2 baryons with light pseudoscalar mesons,''
\href{https://doi.org/10.1088/1674-1137/ac224b}{Chin. Phys. C \textbf{45}, no.11, 113107 (2021)},
\href{https://arxiv.org/abs/2102.03852}{[arXiv:2102.03852 [hep-ph]]}.
%2 citations counted in INSPIRE as of 28 Nov 2023


%\cite{Aliev:2021hqq}
\bibitem{Aliev:2021hqq}
T.~M.~Aliev, T.~Barakat and K.~\c{S}im\c{s}ek,
``Strong $ B_{QQ'}^* B_{QQ'} V $ vertices and the radiative decays of $ B_{QQ}^* \to B_{QQ} \gamma $ in the light-cone sum rules,''
\href{https://doi.org/10.1140/epja/s10050-021-00471-2}{Eur. Phys. J. A \textbf{57}, no.5, 160 (2021)},
\href{https://arxiv.org/abs/2101.10264}{[arXiv:2101.10264 [hep-ph]]}.
%6 citations counted in INSPIRE as of 08 Nov 2023


%\cite{Aliev:2020lly}
\bibitem{Aliev:2020lly}
T.~M.~Aliev and K.~\c{S}im\c{s}ek,
``Strong vertices of doubly heavy spin- $3/2$ \textendash{}spin- $1/2$ baryons with light mesons in light-cone QCD sum rules,''
\href{https://doi:10.1103/PhysRevD.103.054044}{Phys. Rev. D \textbf{103}, no.5, 054044 (2021)},
\href{https://arxiv.org/abs/2011.07150}{[arXiv:2011.07150 [hep-ph]]}.
%5 citations counted in INSPIRE as of 27 Aug 2024 

%\cite{Rostami:2020euc}
\bibitem{Rostami:2020euc}
S.~Rostami, K.~Azizi and A.~R.~Olamaei,
``Strong Coupling Constants of the Doubly Heavy Spin-1/2 Baryons with Light Pseudoscalar Mesons,''
\href{https://doi.org/10.1088/1674-1137/abd084}{Chin. Phys. C \textbf{45}, no.2, 023120 (2021)},
\href{https://arxiv.org/abs/2008.12715}{[arXiv:2008.12715 [hep-ph]]}.
%7 citations counted in INSPIRE as of 28 Nov 2023



%\cite{Olamaei:2020bvw}
\bibitem{Olamaei:2020bvw}
A.~R.~Olamaei, K.~Azizi and S.~Rostami,
``Strong coupling constants of the doubly heavy $ \Xi _{QQ} $ Baryons with $ \pi $ Meson,''
\href{https://doi.org/10.1140/epjc/s10052-020-8194-1}{Eur. Phys. J. C \textbf{80}, no.7, 613 (2020)},
\href{https://arxiv.org/abs/2003.12723}{[arXiv:2003.12723 [hep-ph]]}.
%12 citations counted in INSPIRE as of 08 Nov 2023

%\cite{Aliev:2020aon}
\bibitem{Aliev:2020aon}
T.~M.~Aliev and K.~\c{S}im\c{s}ek,
``Strong coupling constants of doubly heavy baryons with vector mesons in QCD,''
\href{https://doi:10.1140/epjc/s10052-020-08553-z}{Eur. Phys. J. C \textbf{80}, no.10, 976 (2020)},
\href{https://arxiv.org/abs/2009.03464}{[arXiv:2009.03464 [hep-ph]]}.
%11 citations counted in INSPIRE as of 27 Aug 2024




%\cite{Azizi:2020zin}
\bibitem{Azizi:2020zin}
K.~Azizi, A.~R.~Olamaei and S.~Rostami,
``Strong interaction of doubly heavy spin-3/2 baryons with light vector mesons,''
\href{https://doi.org/10.1140/epjc/s10052-020-08770-6}{Eur. Phys. J. C \textbf{80}, no.12, 1196 (2020)},
\href{https://arxiv.org/abs/2011.02919}{[arXiv:2011.02919 [hep-ph]]}.
%6 citations counted in INSPIRE as of 08 Nov 2023

%\cite{Qin:2021dqo}
\bibitem{Qin:2021dqo}
Q.~Qin, Y.~J.~Shi, W.~Wang, G.~H.~Yang, F.~S.~Yu and R.~Zhu,
``Inclusive approach to hunt for the beauty-charmed baryons~$\Xi_{bc}$,''
\href{https://doi.org/10.1103/PhysRevD.105.L031902}{Phys. Rev. D \textbf{105}, no.3, L031902 (2022)},
\href{https://arxiv.org/abs/2108.06716}{[arXiv:2108.06716 [hep-ph]]}.
%25 citations counted in INSPIRE as of 29 Nov 2023






%\cite{Xiao:2017dly}
\bibitem{Xiao:2017dly}
L.~Y.~Xiao, Q.~F.~L\"u and S.~L.~Zhu,
``Strong decays of the 1P and 2D doubly charmed states,''
\href{https://doi.org/10.1103/PhysRevD.97.074005}{Phys. Rev. D \textbf{97}, no.7, 074005 (2018)},
\href{https://arxiv.org/abs/1712.07295}{[arXiv:1712.07295 [hep-ph]]}.
%25 citations counted in INSPIRE as of 08 Nov 2023

%\cite{Xiao:2017udy}
\bibitem{Xiao:2017udy}
L.~Y.~Xiao, K.~L.~Wang, Q.~f.~Lu, X.~H.~Zhong and S.~L.~Zhu,
``Strong and radiative decays of the doubly charmed baryons,''
\href{https://doi.org/10.1103/PhysRevD.96.094005}{Phys. Rev. D \textbf{96}, no.9, 094005 (2017)},
\href{https://arxiv.org/abs/1708.04384}{[arXiv:1708.04384 [hep-ph]]}.
%73 citations counted in INSPIRE as of 22 Nov 2023

%\cite{Li:2017pxa}
\bibitem{Li:2017pxa}
H.~S.~Li, L.~Meng, Z.~W.~Liu and S.~L.~Zhu,
``Radiative decays of the doubly charmed baryons in chiral perturbation theory,''
\href{https://doi.org/10.1016/j.physletb.2017.12.031}{Phys. Lett. B \textbf{777}, 169-176 (2018)},
\href{https://arxiv.org/abs/1708.03620}{[arXiv:1708.03620 [hep-ph]]}.
%69 citations counted in INSPIRE as of 22 Nov 2023




%\cite{Zhao:2018mrg}
\bibitem{Zhao:2018mrg}
Z.~X.~Zhao,
``Weak decays of doubly heavy baryons: the $1/2\rightarrow 3/2$ case,''
\href{https://doi.org/10.1140/epjc/s10052-018-6213-2}{Eur. Phys. J. C \textbf{78}, no.9, 756 (2018)},
\href{https://arxiv.org/abs/1805.10878}{[arXiv:1805.10878 [hep-ph]]}.
%53 citations counted in INSPIRE as of 08 Nov 2023

%\cite{Xing:2018lre}
\bibitem{Xing:2018lre}
Z.~P.~Xing and Z.~X.~Zhao,
``Weak decays of doubly heavy baryons: the FCNC processes,''
\href{https://doi.org/10.1103/PhysRevD.98.056002}{Phys. Rev. D \textbf{98}, no.5, 056002 (2018)},
\href{https://arxiv.org/abs/1807.03101}{[arXiv:1807.03101 [hep-ph]]}.
%39 citations counted in INSPIRE as of 08 Nov 2023

%\cite{Jiang:2018oak}
\bibitem{Jiang:2018oak}
L.~J.~Jiang, B.~He and R.~H.~Li,
``Weak decays of doubly heavy baryons: $\mathcal{B}_{cc}\rightarrow \mathcal{B}_c V$,''
\href{https://doi.org/10.1140/epjc/s10052-018-6445-1}{Eur. Phys. J. C \textbf{78}, no.11, 961 (2018)},
\href{https://arxiv.org/abs/1810.00541}{[arXiv:1810.00541 [hep-ph]]}.
%35 citations counted in INSPIRE as of 08 Nov 2023


%\cite{Gerasimov:2019jwp}
\bibitem{Gerasimov:2019jwp}
A.~S.~Gerasimov and A.~V.~Luchinsky,
``Weak decays of doubly heavy baryons: Decays to a system of $\pi$ mesons,''
\href{https://doi.org/10.1103/PhysRevD.100.073015}{Phys. Rev. D \textbf{100}, no.7, 073015 (2019)},
\href{https://arxiv.org/abs/1905.11740}{[arXiv:1905.11740 [hep-ph]]}.
%19 citations counted in INSPIRE as of 28 Nov 2023

%\cite{Wang:2017mqp}
\bibitem{Wang:2017mqp}
W.~Wang, F.~S.~Yu and Z.~X.~Zhao,
``Weak decays of doubly heavy baryons: the $1/2\rightarrow 1/2$ case,''
\href{https://doi.org/10.1140/epjc/s10052-017-5360-1}{Eur. Phys. J. C \textbf{77}, no.11, 781 (2017)},
\href{https://arxiv.org/abs/1707.02834}{[arXiv:1707.02834 [hep-ph]]}.
%140 citations counted in INSPIRE as of 22 Nov 2023



%\cite{Sharma:2017txj}
\bibitem{Sharma:2017txj}
N.~Sharma and R.~Dhir,
``Estimates of W-exchange contributions to $\Xi_{cc}$ decays,''
\href{https://doi.org/10.1103/PhysRevD.96.113006}{Phys. Rev. D \textbf{96}, no.11, 113006 (2017)},
\href{https://arxiv.org/abs/1709.08217}{[arXiv:1709.08217 [hep-ph]]}.
%44 citations counted in INSPIRE as of 17 Jan 2024





%\cite{Patel:2024mfn}
\bibitem{Patel:2024mfn}
K.~Patel and K.~Thakkar,
``Transition properties of Doubly Heavy Baryons,''
\href{https://arxiv.org/abs/2408.00335}{[arXiv:2408.00335 [hep-ph]]}.
%0 citations counted in INSPIRE as of 16 Aug 2024




%\cite{Gutsche:2019wgu}
\bibitem{Gutsche:2019wgu}
T.~Gutsche, M.~A.~Ivanov, J.~G.~K\"orner and V.~E.~Lyubovitskij,
``Novel ideas in nonleptonic decays of double heavy baryons,''
\href{https://doi.org/10.3390/particles2020021}{Particles \textbf{2}, no.2, 339-356 (2019)},
\href{https://arxiv.org/abs/1905.06219}{[arXiv:1905.06219 [hep-ph]]}.
%16 citations counted in INSPIRE as of 29 Nov 2023



%\cite{Gutsche:2019iac}
\bibitem{Gutsche:2019iac}
T.~Gutsche, M.~A.~Ivanov, J.~G.~K\"orner, V.~E.~Lyubovitskij and Z.~Tyulemissov,
``Analysis of the semileptonic and nonleptonic two-body decays of the double heavy charm baryon states $\Xi_{cc}^{++},\,\Xi_{cc}^{+}$ and $\Omega_{cc}^+$,''
\href{https://doi.org/10.1103/PhysRevD.100.114037}{Phys. Rev. D \textbf{100}, no.11, 114037 (2019)},
\href{https://arxiv.org/abs/1911.10785}{[arXiv:1911.10785 [hep-ph]]}.
%29 citations counted in INSPIRE as of 08 Nov 2023

%\cite{Ke:2019lcf}
\bibitem{Ke:2019lcf}
H.~W.~Ke, F.~Lu, X.~H.~Liu and X.~Q.~Li,
``Study on $\Xi_{cc}\to\Xi_c$ and $\Xi_{cc}\to\Xi'_c$ weak decays in the light-front quark model,''
\href{https://doi.org/10.1140/epjc/s10052-020-7699-y}{Eur. Phys. J. C \textbf{80}, no.2, 140 (2020)},
\href{https://arxiv.org/abs/1912.01435}{[arXiv:1912.01435 [hep-ph]]}.
%24 citations counted in INSPIRE as of 08 Nov 2023

%\cite{Cheng:2020wmk}
\bibitem{Cheng:2020wmk}
H.~Y.~Cheng, G.~Meng, F.~Xu and J.~Zou,
``Two-body weak decays of doubly charmed baryons,''
\href{https://doi.org/10.1103/PhysRevD.101.034034}{Phys. Rev. D \textbf{101}, no.3, 034034 (2020)},
\href{https://arxiv.org/abs/2001.04553}{[arXiv:2001.04553 [hep-ph]]}.
%26 citations counted in INSPIRE as of 08 Nov 2023

%\cite{Hu:2020mxk}
\bibitem{Hu:2020mxk}
X.~H.~Hu, R.~H.~Li and Z.~P.~Xing,
``A comprehensive analysis of weak transition form factors for doubly heavy baryons in the light front approach,''
\href{https://doi.org/10.1140/epjc/s10052-020-7851-8}{Eur. Phys. J. C \textbf{80}, no.4, 320 (2020)},
\href{https://arxiv.org/abs/2001.06375}{[arXiv:2001.06375 [hep-ph]]}.
%31 citations counted in INSPIRE as of 08 Nov 2023



%\cite{Li:2020qrh}
\bibitem{Li:2020qrh}
R.~H.~Li, J.~J.~Hou, B.~He and Y.~R.~Wang,
``Weak Decays of Doubly Heavy Baryons: ${\cal B}_{cc}\to {\cal B} D^{(*)}$,''
\href{https://doi.org/10.1088/1674-1137/abe0bc}{Chinese Phys. C 45 043108 (2021)},
\href{https://arxiv.org/abs/2010.09362}{[arXiv:2010.09362 [hep-ph]]}.
%8 citations counted in INSPIRE as of 08 Nov 2023

%\cite{Han:2021gkl}
\bibitem{Han:2021gkl}
J.~J.~Han, R.~X.~Zhang, H.~Y.~Jiang, Z.~J.~Xiao and F.~S.~Yu,
``Weak decays of bottom-charm baryons: $\mathcal {B}_{bc}\rightarrow \mathcal {B}_bP$,''
\href{https://doi.org/10.1140/epjc/s10052-021-09239-w}{Eur. Phys. J. C \textbf{81}, no.6, 539 (2021)},
\href{https://arxiv.org/abs/2102.00961}{[arXiv:2102.00961 [hep-ph]]}.
%12 citations counted in INSPIRE as of 08 Nov 2023

%\cite{Wang:2017azm}
\bibitem{Wang:2017azm}
W.~Wang, Z.~P.~Xing and J.~Xu,
``Weak Decays of Doubly Heavy Baryons: SU(3) Analysis,''
\href{https://doi.org/10.1140/epjc/s10052-017-5363-y}{Eur. Phys. J. C \textbf{77}, no.11, 800 (2017)},
\href{https://arxiv.org/abs/1707.06570}{[arXiv:1707.06570 [hep-ph]]}.
%112 citations counted in INSPIRE as of 28 Nov 2023

%\cite{Shi:2017dto}
\bibitem{Shi:2017dto}
Y.~J.~Shi, W.~Wang, Y.~Xing and J.~Xu,
``Weak Decays of Doubly Heavy Baryons: Multi-body Decay Channels,''
\href{https://doi.org/10.1140/epjc/s10052-018-5532-7}{Eur. Phys. J. C \textbf{78}, no.1, 56 (2018)},
\href{https://arxiv.org/abs/1712.03830}{[arXiv:1712.03830 [hep-ph]]}.
%70 citations counted in INSPIRE as of 22 Nov 2023





%\cite{Ivanov:2020xmw}
\bibitem{Ivanov:2020xmw}
M.~A.~Ivanov, J.~G.~K\"orner and V.~E.~Lyubovitskij,
``Nonleptonic Decays of Doubly Charmed Baryons,''
\href{https://doi.org/10.1134/S1063779620040358}{Phys. Part. Nucl. \textbf{51}, no.4, 678-685 (2020)},
%10 citations counted in INSPIRE as of 08 Nov 2023



%\cite{Shi:2020qde}
\bibitem{Shi:2020qde}
Y.~J.~Shi, W.~Wang, Z.~X.~Zhao and U.~G.~Mei\ss{}ner,
``Towards a Heavy Diquark Effective Theory for Weak Decays of Doubly Heavy Baryons,''
\href{https://doi.org/10.1140/epjc/s10052-020-7949-z}{Eur. Phys. J. C \textbf{80}, no.5, 398 (2020)},
\href{https://arxiv.org/abs/2002.02785}{[arXiv:2002.02785 [hep-ph]]}.
%20 citations counted in INSPIRE as of 29 Nov 2023





%\cite{Hu:2017dzi}
\bibitem{Hu:2017dzi}
X.~H.~Hu, Y.~L.~Shen, W.~Wang and Z.~X.~Zhao,
``Weak decays of doubly heavy baryons: ''decay constants'',''
\href{https://doi.org/10.1088/1674-1137/42/12/123102}{Chin. Phys. C \textbf{42}, no.12, 123102 (2018)},
\href{https://arxiv.org/abs/1711.10289}{[arXiv:1711.10289 [hep-ph]]}.
%39 citations counted in INSPIRE as of 08 Nov 2023

%\cite{Li:2018epz}
\bibitem{Li:2018epz}
R.~H.~Li and C.~D.~Lu,
``Search for doubly heavy baryon via weak decays,''
\href{https://arxiv.org/abs/1805.09064}{[arXiv:1805.09064 [hep-ph]]}.
%8 citations counted in INSPIRE as of 08 Nov 2023

%\cite{Shi:2019hbf}
\bibitem{Shi:2019hbf}
Y.~J.~Shi, W.~Wang and Z.~X.~Zhao,
``QCD Sum Rules Analysis of Weak Decays of Doubly-Heavy Baryons,''
\href{https://doi.org/10.1140/epjc/s10052-020-8096-2}{Eur. Phys. J. C \textbf{80}, no.6, 568 (2020)},
\href{https://arxiv.org/abs/1902.01092}{[arXiv:1902.01092 [hep-ph]]}.
%51 citations counted in INSPIRE as of 08 Nov 2023



%\cite{Shi:2019fph}
\bibitem{Shi:2019fph}
Y.~J.~Shi, Y.~Xing and Z.~X.~Zhao,
``Light-cone sum rules analysis of $\Xi_{QQ^{\prime}q}\to\Lambda_{Q^{\prime}}$ weak decays,''
\href{https://doi.org/10.1140/epjc/s10052-019-7014-y}{Eur. Phys. J. C \textbf{79}, no.6, 501 (2019)},
\href{https://arxiv.org/abs/1903.03921}{[arXiv:1903.03921 [hep-ph]]}.
%38 citations counted in INSPIRE as of 22 Nov 2023

%\cite{Zhang:2018llc}
\bibitem{Zhang:2018llc}
Q.~A.~Zhang,
``Weak Decays of Doubly Heavy Baryons: W-Exchange,''
\href{https://doi.org/10.1140/epjc/s10052-018-6481-x}{Eur. Phys. J. C \textbf{78}, no.12, 1024 (2018)},
\href{https://arxiv.org/abs/1811.02199}{[arXiv:1811.02199 [hep-ph]]}.
%28 citations counted in INSPIRE as of 08 Nov 2023


%\cite{Gutsche:2017hux}
\bibitem{Gutsche:2017hux}
T.~Gutsche, M.~A.~Ivanov, J.~G.~K\"orner and V.~E.~Lyubovitskij,
``Decay chain information on the newly discovered double charm baryon state $\Xi_{cc}^{++}$,''
\href{https://doi.org/10.1103/PhysRevD.96.054013}{Phys. Rev. D \textbf{96}, no.5, 054013 (2017)},
\href{https://arxiv.org/abs/1708.00703}{[arXiv:1708.00703 [hep-ph]]}.
%55 citations counted in INSPIRE as of 08 Nov 2023

%\cite{Gutsche:2018msz}
\bibitem{Gutsche:2018msz}
T.~Gutsche, M.~A.~Ivanov, J.~G.~K\"orner, V.~E.~Lyubovitskij and Z.~Tyulemissov,
``Ab initio three-loop calculation of the $W$-exchange contribution to nonleptonic decays of double charm baryons,''
\href{https://doi.org/10.1103/PhysRevD.99.056013}{Phys. Rev. D \textbf{99}, no.5, 056013 (2019)},
\href{https://arxiv.org/abs/1812.09212}{[arXiv:1812.09212 [hep-ph]]}.
%36 citations counted in INSPIRE as of 08 Nov 2023



%\cite{Ozdem:2018uue}
\bibitem{Ozdem:2018uue}
U.~\"Ozdem,
``Magnetic moments of doubly heavy baryons in light-cone QCD,''
\href{https://doi.org/10.1088/1361-6471/aafffc}{J. Phys. G \textbf{46}, no.3, 035003 (2019)},
\href{https://arxiv.org/abs/1804.10921}{[arXiv:1804.10921 [hep-ph]]}.

%\cite{Ozdem:2019zis}
\bibitem{Ozdem:2019zis}
U.~\"Ozdem,
``Magnetic dipole moments of the spin-$\frac{3}{2}$ doubly heavy baryons,''
\href{https://doi.org/10.1140/epja/s10050-020-00049-4}{Eur. Phys. J. A \textbf{56}, no.2, 34 (2020)},
\href{https://arxiv.org/abs/1906.08353}{[arXiv:1906.08353 [hep-ph]]}.



%\cite{Berezhnoy:2018bde}
\bibitem{Berezhnoy:2018bde}
A.~V.~Berezhnoy, A.~K.~Likhoded and A.~V.~Luchinsky,
``Doubly heavy baryons at the LHC,''
\href{https://doi.org/10.1103/PhysRevD.98.113004}{Phys. Rev. D \textbf{98}, no.11, 113004 (2018)},
\href{https://arxiv.org/abs/1809.10058}{[arXiv:1809.10058 [hep-ph]]}.
%41 citations counted in INSPIRE as of 08 Nov 2023




%\cite{Shifman:1978bx}
\bibitem{Shifman:1978bx}
M.~A.~Shifman, A.~I.~Vainshtein and V.~I.~Zakharov,
``QCD and Resonance Physics. Theoretical Foundations,''
\href{https://www.sciencedirect.com/science/article/abs/pii/0550321379900221}{Nucl. Phys. B \textbf{147}, 385-447 (1979)}.
%5730 citations counted in INSPIRE as of 01 Dec 2023

%\cite{Shifman:1978by}
\bibitem{Shifman:1978by}
M.~A.~Shifman, A.~I.~Vainshtein and V.~I.~Zakharov,
``QCD and Resonance Physics: Applications,''
\href{https://www.sciencedirect.com/science/article/abs/pii/0550321379900233}{Nucl. Phys. B \textbf{147}, 448-518 (1979)}.
%3144 citations counted in INSPIRE as of 23 Nov 2023

%\cite{Aliev:2010uy}
\bibitem{Aliev:2010uy}
T.~M.~Aliev, K.~Azizi and M.~Savci,
``Analysis of the $\Lambda_{b}\rightarrow \Lambda \ell^+\ell^- $ decay in QCD,''
\href{https://doi.org/10.1103/PhysRevD.81.056006}{Phys. Rev. D \textbf{81}, 056006 (2010)},
\href{https://arxiv.org/abs/1001.0227}{[arXiv:1001.0227 [hep-ph]]}.

%\cite{Aliev:2009jt}
\bibitem{Aliev:2009jt}
T.~M.~Aliev, K.~Azizi and A.~Ozpineci,
``Radiative Decays of the Heavy Flavored Baryons in Light Cone QCD Sum Rules,''
\href{https://doi.org/10.1103/PhysRevD.79.056005}{Phys. Rev. D \textbf{79}, 056005 (2009)},
\href{https://arxiv.org/abs/0901.0076}{[arXiv:0901.0076 [hep-ph]]}.



%\cite{Agaev:2016dev}
\bibitem{Agaev:2016dev}
S.~S.~Agaev, K.~Azizi and H.~Sundu,
``Strong $Z_c^{+}(3900)\rightarrow J/\psi \pi^{+}; \eta_{c} \rho^{+}$ decays in QCD,''
\href{https://doi.org/10.1103/PhysRevD.93.074002}{Phys. Rev. D \textbf{93}, no.7, 074002 (2016)},
\href{https://arxiv.org/abs/1601.03847}{[arXiv:1601.03847 [hep-ph]]}.

%\cite{Azizi:2016dhy}
\bibitem{Azizi:2016dhy}
K.~Azizi, Y.~Sarac and H.~Sundu,
``Analysis of $P_c^+(4380)$ and $P_c^+(4450)$ as pentaquark states in the molecular picture with QCD sum rules,''
\href{https://doi.org/10.1103/PhysRevD.95.094016}{Phys. Rev. D \textbf{95}, no.9, 094016 (2017)},
\href{https://arxiv.org/abs/1612.07479}{[arXiv:1612.07479 [hep-ph]]}.

%\cite{Wang:2007ys}
\bibitem{Wang:2007ys}
Y.~M.~Wang, H.~Zou, Z.~T.~Wei, X.~Q.~Li and C.~D.~Lu,
``The Transition form-factors for semi-leptonic weak decays of J / psi in QCD sum rules,''
\href{https://doi:10.1140/epjc/s10052-007-0498-x}{Eur. Phys. J. C \textbf{54}, 107-121 (2008)},
\href{https://arxiv.org/abs/0707.1138}{[arXiv:0707.1138 [hep-ph]]}.
%49 citations counted in INSPIRE as of 01 Oct 2024

%\cite{Azizi:2018axf}
\bibitem{Azizi:2018axf}
K.~Azizi and J.~Y.~S\"ung\"u,
``Semileptonic $\Lambda_{b}\rightarrow \Lambda_{c}{\ell}\bar\nu_{\ell}$ Transition in Full QCD,''
\href{https://doi:10.1103/PhysRevD.97.074007}{Phys. Rev. D \textbf{97}, no.7, 074007 (2018)},
\href{https://arxiv.org/abs/1803.02085}{[arXiv:1803.02085 [hep-ph]]}.
%31 citations counted in INSPIRE as of 20 Jun 2024

%\cite{Aliev:2006gk}
\bibitem{Aliev:2006gk}
T.~M.~Aliev, K.~Azizi and A.~Ozpineci,
``Semileptonic B(s) ---\ensuremath{>} D(sJ)(2460)l nu decay in QCD,''
\href{https://doi:10.1140/epjc/s10052-007-0315-6}{Eur. Phys. J. C \textbf{51}, 593-599 (2007)}'
\href{https://arxiv.org/abs/hep-ph/0608264}{[arXiv:hep-ph/0608264 [hep-ph]]}.
%31 citations counted in INSPIRE as of 14 May 2024


%\cite{Aliev:2010yx}
\bibitem{Aliev:2010yx}
T.~M.~Aliev, K.~Azizi and M.~Savci,
``Strong coupling constants of light pseudoscalar mesons with heavy baryons in QCD,''
\href{https://doi:10.1016/j.physletb.2010.12.027}{Phys. Lett. B \textbf{696}, 220-226 (2011)},
\href{https://arxiv.org/abs/hep-ph/1009.3658}{[arXiv:1009.3658 [hep-ph]]}.
%34 citations counted in INSPIRE as of 06 Nov 2024

%\cite{Agaev:2020zad}
\bibitem{Agaev:2020zad}
S.~Agaev, K.~Azizi and H.~Sundu,
``Four-quark exotic mesons,''
\href{https://doi:10.3906/fiz-2003-15}{Turk. J. Phys. \textbf{44}, no.2, 95-173 (2020)},
\href{https://arxiv.org/abs/2004.12079}{[arXiv:2004.12079 [hep-ph]]}.
%68 citations counted in INSPIRE as of 13 Jun 2024


%\cite{Zyla:2020zbs}
\bibitem{ParticleDataGroup:2020ssz}
P.~A.~Zyla \textit{et al.} [Particle Data Group],
``Review of Particle Physics,''
\href{https://academic.oup.com/ptep/article/2020/8/083C01/5891211?login=false}{PTEP \textbf{2020}, no.8, 083C01 (2020)}.
%doi:10.1093/ptep/ptaa104
%5535 citations counted in INSPIRE as of 01 Dec 2023

%\cite{Belyaev:1982sa}
\bibitem{Belyaev:1982sa}
V.~M.~Belyaev and B.~L.~Ioffe,
``Determination of Baryon and Baryonic Resonance Masses from QCD Sum Rules. 1. Nonstrange Baryons,''
\href{http://www.jetp.ras.ru/cgi-bin/e/index/e/56/3/p493?a=list}{Sov. Phys. JETP \textbf{56}, 493-501 (1982)}
ITEP-59-1982.
%416 citations counted in INSPIRE as of 08 Nov 2023

%\cite{Belyaev:1982cd}
\bibitem{Belyaev:1982cd}
V.~M.~Belyaev and B.~L.~Ioffe,
``Determination of the baryon mass and baryon resonances from the quantum-chromodynamics sum rule. Strange baryons,''
\href{http://www.jetp.ras.ru/cgi-bin/e/index/e/57/4/p716?a=list}{Sov. Phys. JETP \textbf{57}, 716-721 (1983)}
ITEP-132-1982.
%180 citations counted in INSPIRE as of 08 Nov 2023
%\cite{Ioffe:2005ym}

%\cite{Ioffe:2005ym}
\bibitem{Ioffe:2005ym}
B.~L.~Ioffe,
"QCD at low energies,''
\href{https://doi.org/10.1016/j.ppnp.2005.05.001}{Prog. Part. Nucl. Phys. \textbf{56}, 232-277 (2006)},
\href{https://arxiv.org/abs/hep-ph/0502148.pdf}{[arXiv:hep-ph/0502148 [hep-ph]]}.
%408 citations counted in INSPIRE as of 31 Jan 2024

%\cite{Wang:2010fq}
\bibitem{Wang:2010fq}
Z.~G.~Wang,
``Analysis of the ${1\over 2}^{\pm}$ antitriplet heavy baryon states with QCD sum rules,''
\href{https://link.springer.com/article/10.1140/epjc/s10052-010-1365-8}{Eur. Phys. J. C \textbf{68}, 479-486 (2010)},
\href{https://arxiv.org/pdf/1001.1652}{[arXiv:1001.1652 [hep-ph]]}.
%43 citations counted in INSPIRE as of 04 Jun 2024

%\cite{Wang:2015vgv}
\bibitem{Wang:2015vgv}
Y.~M.~Wang and Y.~L.~Shen,
``QCD corrections to B\textrightarrow{}\ensuremath{\pi} form factors from light-cone sum rules,''
\href{https://doi:10.1016/j.nuclphysb.2015.07.016}{Nucl. Phys. B \textbf{898}, 563-604 (2015)},
\href{https://arxiv.org/pdf/1506.00667}{[arXiv:1506.00667 [hep-ph]]}.
%93 citations counted in INSPIRE as of 01 Oct 2024

%\cite{Cui:2022zwm}
\bibitem{Cui:2022zwm}
B.~Y.~Cui, Y.~K.~Huang, Y.~L.~Shen, C.~Wang and Y.~M.~Wang,
``Precision calculations of B$_{d,s}$ \textrightarrow{} \ensuremath{\pi}, K decay form factors in soft-collinear effective theory,''
\href{https://doi:10.1007/JHEP03(2023)140}{JHEP \textbf{03}, 140 (2023)},
\href{https://arxiv.org/pdf/2212.11624}{[arXiv:2212.11624 [hep-ph]]}.
%25 citations counted in INSPIRE as of 01 Oct 2024


%\cite{Khodjamirian:2011jp}
\bibitem{Khodjamirian:2011jp}
A.~Khodjamirian, C.~Klein, T.~Mannel and Y.~M.~Wang,
``Form Factors and Strong Couplings of Heavy Baryons from QCD Light-Cone Sum Rules,''
\href{https://doi:10.1007/JHEP09(2011)106}{JHEP \textbf{09}, 106 (2011)},
\href{https://arxiv.org/pdf/1108.2971}{[arXiv:1108.2971 [hep-ph]]}.
%135 citations counted in INSPIRE as of 01 Oct 2024


\end{thebibliography}
\end{document}